\documentclass[10pt,a4paper,english,pra, twocolumn]{revtex4-1}
\usepackage[T1]{fontenc}
\usepackage[latin9]{inputenc}
\usepackage{color}
\definecolor{page_backgroundcolor}{rgb}{1, 1, 1}
\pagecolor{page_backgroundcolor}
\usepackage{float}
\usepackage{amsmath}
\usepackage{amssymb}
\usepackage{stmaryrd}
\usepackage{graphicx}

\makeatletter

\pdfpageheight\paperheight
\pdfpagewidth\paperwidth

\usepackage[colorlinks,
            linkcolor=blue,
            anchorcolor=blue,
            citecolor=blue
            ]{hyperref}

\makeatother

\usepackage{babel}
\begin{document}
\title{Non-Gaussian Quantum State Engineering with Postselected von Neumann
Measurements}
\author{Xiao-Xi Yao }
\author{Yusuf Turek}
\email{Corresponding author: yusuftu1984@hotmail.com}

\affiliation{$^{1}$School of Physics, Liaoning University,Shenyang,Liaoning 110036,China}
\date{\today}
\begin{abstract}
We introduce a feasible protocol for generating non-Gaussian (nG)
states via postselected von Neumann measurement for continuous-variable
quantum information processing. The method uses a two-level system
coupled to a Gaussian pointer state through an observable $A$ with
$A^{2}=\mathbb{I}$. By operating beyond the weak-coupling regime
and selecting different pointer states-{}-squeezed, coherent, or vacuum-{}-allows
for the generation of a wide range of nG states, including squeezed
cat states, two-mode entangled cat states, approximate Bell states,
and a continuum of intermediate nG states with considerable success
probabilities. The properties of these states are widely tunable via
the postselection-induced weak value and the measurement interaction
strength. We characterize the non-Gaussianity via Wigner function
negativities and quantify entanglement using linear entropy and concurrence.
The protocol offers a scalable route to high-purity nG state engineering. 
\end{abstract}
\maketitle

\section{Introduction}

Non-Gaussian (nG) states \citep{PRXQuantum.2.030204,PRXQuantum.1.020305},
owing to their intrinsic Wigner function negativity and higher-order
quantum correlations, constitute essential resources for overcoming
various limitations within the Gaussian framework. They have demonstrated
great potential in quantum computation \citep{PhysRevLett.82.1784,PhysRevLett.109.230503,PhysRevA.93.022301,PhysRevLett.97.110501,PhysRevA.64.012310},
quantum metrology \citep{PhysRevLett.104.103602,PhysRevA.78.063828},
and foundational studies in quantum theory \citep{PhysRevA.67.012106,PhysRevA.75.052105}.
However, the generation of high-quality nG states is a crucial yet
challenging task in quantum resource engineering. The generation of
nG states is commonly achieved with conditional measurements including
photon addition/subtraction \citep{2004,doi:10.1126/science.1146204,Wakui:07,PhysRevA.110.023703},
photon catalysis \citep{PhysRevLett.88.250401,PhysRevLett.96.083601},
and nonlinear optical processes \citep{PhysRevLett.57.13,PhysRevA.55.2478}.
Although nonlinear optical processes can, in principle, generate nG
states directly, their practical implementation is hindered by weak
nonlinearities, significant loss and noise, limited controllability
of the evolution and poor scalability. As a result, more feasible
experimental schemes rely on nG operation-based conditional measurements
applied to specific given states. As a paradigmatic example of nG
states, Schr�dinger cat states have been realized through various
conditional measurement schemes. A well-known approach is based on
adding or subtracting a single photon from a squeezed vacuum state;
however, this method is restricted to the preparation of small-amplitude
cat states ($\alpha\le1.2$) \citep{2006}. From the perspective of
both fundamental quantum theory and practical applications in quantum
information processing, large-amplitude cat states ($\alpha\geq2$)
are of greater significance \citep{PhysRevA.64.022313,PhysRevA.69.022315,PhysRevA.68.022321}.
Another prominent approach employs conditional measurements on Fock
states to generate squeezed cat states with arbitrary coherent amplitudes
\citep{ourjoumtsev2007generation}. While this method has successfully
demonstrated the preparation of large-amplitude cat states (with $\alpha>2$),
it requires costly quantum resources and suffers from very low success
probabilities, which severely limit its practicality and scalability. 

Furthermore, in recent years, increasing attention has been devoted
to the preparation of nG states \citep{PhysRevA.105.022608,RN11,yao2025nongaussianstatepreparationenhancement}
via weak value amplification effects of weak measurements \citep{PhysRevLett.60.1351,Nature2017}.
Under weak-coupling conditions, the von Neumann--type Hamiltonian
can be approximated as an effective photon addition/subtraction operation.
By appropriately choosing the pointer state, this method has been
successfully applied to generate various nG states, such as photon-added
coherent states and squeezed number states without actual photon addition/subtraction
operations. However, the success probability of this method remains
sufficiently low due to the restriction of the weak-coupling condition.
To address the above challenges, it is highly desirable to develop
an efficient approach capable of generating high-fidelity nG states
with improved success rates.

In this work, we propose a feasible protocol for nG state engineering
based on postselected von Neumann measurements. The protocol utilizes
a two-level system, pre- and post-selected to yield a weak value of
an observable $A$ satisfying $A^{2}=\mathbb{I}$, coupled to a Gaussian
pointer state. We abandon the approximation inherent in the conventional
weak measurement model and directly investigate the dynamics of a
von Neumann--type interaction Hamiltonian at arbitrary coupling strengths.
Within this framework, the pointer state undergoes displacement evolution
under the action of the Hamiltonian, and postselected measurements
give rise to weak values. By precisely tuning the coupling strength
between the measured system and the pointer, the pointer state can
be steered into a variety of nG states. By eliminating the restriction
of the weak-coupling approximation, this scheme overcomes the intrinsic
bottleneck of low success probability in traditional weak-measurement-based
models and provides a new route toward efficient quantum state engineering.
Specifically, by initializing the pointer state as a squeezed vacuum
state produces squeezed cat-like states, which can be tuned to approximate
Gottesman-Kitaev-Preskill (GKP) states \citep{PhysRevA.64.012310}.
A coherent state input yields superpositions of displaced coherent
states, encompassing odd, even, and Yurke-Stoler-type cat states.
For two-mode scenarios, feeding separable squeezed or coherent states
into the measurement setup, followed by a beam splitter, deterministically
produces two-mode entangled nG states, including squeezed entangled
cat states and Bell-like states approaching maximal entanglement,
and other continuum of intermediate nG states. Crucially, by operating
outside the weak-coupling regime, our method achieves a significantly
higher success probability than photon addition or photon subtraction
based approaches. The non-Gaussianity features of the generated states
are quantitatively characterized using the Wigner function, while
the entanglement strength of the two-mode states is further evaluated
through linear entropy and concurrence.

The remainder of this paper is organized as follows. In Sec. \ref{sec:2},
we introduce our protocol for generating various nG states via a von
Neumann--type measurement. In Secs. \ref{sec:3} and \ref{sec:4},
as examples, we present the details of the state generation for the
single-mode and two-mode input cases, respectively, and provide analyses
of several representative nG states. We also investigate the Wigner
function and entanglement measures of output states. The conclusion
and outlook are given in Sec. \ref{sec:5}. Throughout this paper,
we adopt the natural unit system with $\hbar=1$. The numerical plots
in this work were performed using the Python package QuTiP \citep{JOHANSSON20121760,JOHANSSON20131234}.

\section{\label{sec:2} Model setup }

In this section, we outline our protocol for state preparation using
a postselected von Neumann measurement. Any single-mode or multi-mode
radiation field possesses internal degrees of freedom (e.g., polarization,
spin, energy level) and external degrees of freedom (e.g., spatial,
angular, or temporal distribution). By designating the internal degrees
as the measured system and the external degrees as the pointer (the
measurement apparatus), the interaction Hamiltonian takes the standard
von Neumann form

\begin{equation}
H_{int}=g_{0}A\otimes Q,\label{eq:(1)}
\end{equation}
where $g_{0}$ represents the interaction coupling between the pointer
and the measured system. The observable $A$ acting on the two-level
measured system, is postulated to satisfy $A^{2}=\mathbb{I}$, thereby
constraining its eigenvalues to the range $[-1,1]$. The pointer observable
$Q$ is defined as a canonical quadrature, either position $X$ or
momentum $P$. In terms of the annihilation and creation operators
$a$ and $a^{\dagger}$, we can express $X$ and $P$ as
\begin{equation}
X=\sigma(a+a^{\dagger}),\label{eq:2-1}
\end{equation}
and
\begin{equation}
P=\frac{i}{2\sigma}(a^{\dagger}-a),\label{eq:3-1}
\end{equation}
respectively, with $\sigma$ parameterizing the spatial width of the
pointer state. 

\begin{figure}[H]
\centering
\includegraphics[width=8cm]{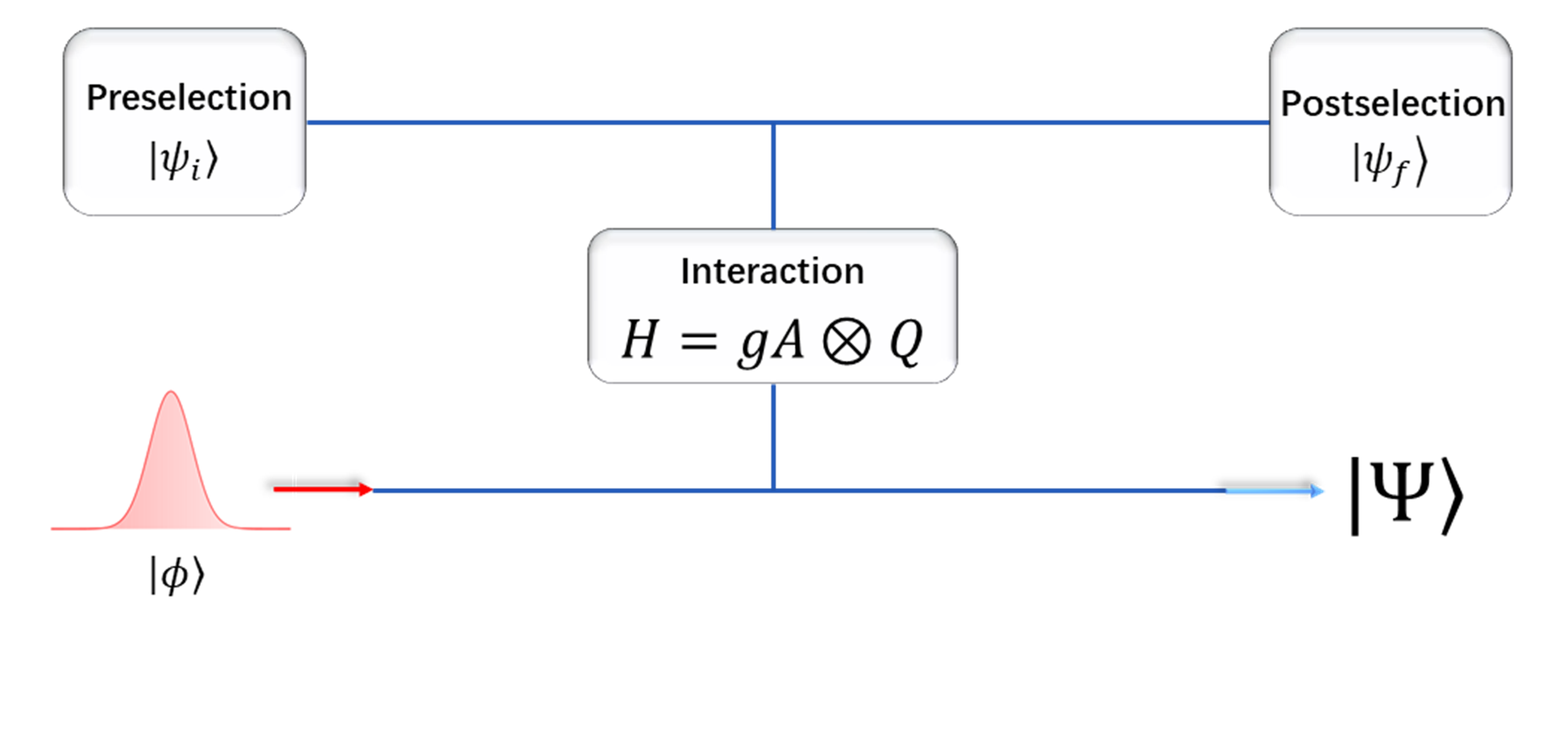}

\caption{\label{fig:1}Schematic diagram of nG state generation via postselected
von Neumann measurement. The pre-selected measured system state and
pointer evolved under interaction coupling and followed by post-selection
implemented on measured system. The output state $\vert\Phi\rangle$
is generated with definite probability.}
\end{figure}

We consider a two-level system initially prepared in state $\vert\psi_{i}\rangle$,
and the pointer in the state $\vert\phi\rangle$, defining the initial
composite system state as $|\varPsi_{in}\rangle=|\psi_{i}\rangle\otimes|\phi\rangle$.
The time evolution governed by the interaction Hamiltonian $H_{int}$
is described by the unitary operator
\begin{align}
U(t) & =exp\left(-i\int_{0}^{t}H_{int}d\tau\right)\nonumber \\
 & =\frac{1}{2}\left[\left(1+A\right)\otimes e^{-igQ}+\left(1-A\right)\otimes e^{igQ}\right],\label{eq:5}
\end{align}
where $g=g_{0}t$. Application of this unitary evolution to the initial
state yields the evolved composite state: 

\begin{equation}
\begin{aligned}\left|\Psi_{evol}\right\rangle  & =\exp\left(-i\int_{0}^{t}H_{int}d\tau\right)|\Psi_{in}\rangle\\
 & =\frac{1}{2}\left[\left(1+A\right)\otimes e^{-igQ}+\left(1-A\right)\otimes e^{igQ}\right]|\psi_{i}\rangle\otimes|\phi\rangle.
\end{aligned}
\label{eq:4}
\end{equation}
 Subsequent postselection of the system onto a final state $|\psi_{f}\rangle$
projects the pointer onto the unnormalized state $\vert\Phi\rangle=\text{\ensuremath{\langle\psi_{f}\vert\Psi_{evol}\rangle}}$.
The final, normalized pointer state is therefore given by 
\begin{align}
\vert\Psi\rangle & =\frac{\vert\Phi\rangle}{\sqrt{p}}= & \frac{1}{2}\frac{\langle\psi_{f}|\psi_{i}\rangle}{\sqrt{p}}\left[t_{+}e^{-igQ}+t_{-}e^{igQ}\right]|\phi\rangle,\label{eq:6}
\end{align}
where the coefficients are defined as $t_{\pm}=1\pm\langle A\rangle_{\omega}$.
Here, $p=\langle\Phi\vert\Phi\rangle$ denotes the successful postselection
probability, and $\langle A\rangle_{w}$ is the weak value of the
observable $A$, defined by the expression
\begin{equation}
\langle A\rangle_{w}=\frac{\langle\psi_{f}|A|\psi_{i}\rangle}{\langle\psi_{f}|\psi_{i}\rangle}.\label{eq:8-2}
\end{equation}

Generally, the weak value $\langle A\rangle_{w}$ is complex-valued
and may lie outside the eigenvalue spectrum of the observable $A$.
It is crucial to note that the derivation of Eq. (\ref{eq:6}) does
not rely on any approximation, and the interaction coupling parameter
$g$ can assume arbitrary values. A schematic of the state-generation
protocol is presented in Fig. \ref{fig:1}. 

The weak value emerges fundamentally from the pre- and post-selection
procedure, with its definition independent of the measurement strength
\citep{PhysRevA.98.042112}. This establishes the postselected von
Neumann measurement as a generalization of the standard weak measurement
formalism, which is typically restricted to a low-order Taylor expansion
of the evolution operator \citep{PhysRevLett.60.1351}. As demonstrated
in previous studies \citep{PhysRevA.83.052106,PhysRevA.93.032128,PhysRevA.98.042112,PhysRevA.105.042202},
the accessibility of weak values is not confined to the weak-measurement
regime; they remain well-defined within the two-state vector formalism
(TSVF) even under strong measurement conditions \citep{Aharonov1964TIMESI,Aharonov2008TheTV}.
Consequently, the weak value constitutes not merely an experimental
artifact of weak interactions but a fundamental element of the operator
algebra \citep{Hofmann_2012,Wagner2024}. 

The properties of the normalized postmeasurement pointer state $\vert\Psi\rangle$
are determined by both the weak value $\langle A\rangle_{w}$ and
the initial input state $\vert\phi\rangle$. As previously noted,
the pointer variable $Q$ can represent either the position or momentum
operator. In this work, we specify $Q=P$, allowing the state $\vert\Psi\rangle$
to be reformulated as 
\begin{equation}
\vert\Psi\rangle=\frac{\vert\Phi\rangle}{\sqrt{p}}=\frac{\langle\psi_{f}|\psi_{i}\rangle}{2\sqrt{p}}\left[t_{+}D\left(s\right)+t_{-}D\left(-s\right)\right]|\phi\rangle,\label{eq:8-1}
\end{equation}
where $D\left(s\right)=e^{s\left(a^{\dagger}-a\right)}$ is the displacement
operator and $s=g/2\sigma$. The regimes $s\ll1$ and $s\gg1$ correspond
to weak and strong measurement couplings, respectively. We therefore
designate $s$ as the interaction strength parameter.

The success probability of the final state $\vert\Psi\rangle$, which
incorporates the postselection probability $\vert\langle\psi_{f}\vert\psi_{i}\rangle\vert^{2}$,
serves as a crucial metric for evaluating the efficiency of our protocol.
While standard postselected weak measurements achieving weak value
amplification through anomalous weak values typically suffer from
low postselection probabilities, we demonstrate that the overall success
probability $p$ remains non-negligible even for substantially anomalous
weak values.

To illustrate this concretely, we consider a specific configuration
where the system observable is $A=\sigma_{x}=\vert V\rangle\langle H\vert+\vert H\rangle\langle V\vert$
, with pre- and post-selected states given by $\vert\psi_{i}\rangle=\cos\frac{\theta}{2}\vert H\rangle+e^{i\delta}\sin\frac{\theta}{2}\vert V\rangle$
and $\vert\psi_{f}\rangle=\vert H\rangle,$ respectively. Here, $\vert H\rangle$
and $\vert V\rangle$denote horizontal and vertical polarization of
a given optical beam, with parameters $\theta\in[0,\pi)$ and $\delta\in[0,2\pi].$
The corresponding weak value is $\langle\sigma_{x}\rangle_{w}=e^{i\delta}\tan\frac{\theta}{2}$,
obtained with a postselection probability of $\cos^{2}\frac{\theta}{2}$.
We use this example to specify the $p$ in concrete examples in next
sections.

The state $\vert\Psi\rangle$ constitutes the final pointer state
following the postselected von Neumann measurement. In subsequent
sections, we explore state preparation protocols for various initial
pointer states associated with single- and two-mode radiation fields.
Notably, all generated states exhibit nG characteristics except in
the special cases where $\langle A\rangle_{w}=\pm1$.

\section{\label{sec:3}Generation of cat-like states }

This section presents the preparation of single-mode cat-like states
using our protocol based on postselected von Neumann measurements.
We demonstrate how superpositions of coherent states---fundamental
nonclassical resources---can be generated and controlled through
appropriate selection of input pointer states and measurement parameters.
The flexibility of our approach allows for the generation of a wide
range of cat-like states, including (squeezed) even, odd, and Yurke--Stoler
type states, with tunable amplitudes and nG characteristics governed
by the weak value and interaction strength. 

\subsection{\label{subsec:A}Squeezed Cat-like states}

The squeezed coherent state represents a fundamental resource in quantum
optics due to its unique properties and applications across quantum
sensing \citep{lawrie2019quantum} and quantum communications \citep{PhysRevA.63.022309,PhysRevA.63.052311},
among other domains \citep{Andersen2016,SCHNABEL20171}. This state,
defined as $\vert\alpha,\xi\rangle=D(\alpha)S(\xi)\vert0\rangle$,
is generated by applying the displacement operator $D(\alpha)=$$\exp\left[\alpha a^{\dag}-\alpha^{*}a\right]$
and squeezing operator $S(\xi)=\exp\left[\frac{\xi^{\ast}}{2}a^{2}-\frac{\xi}{2}a^{\dagger2}\right]$
to the vacuum state, where $\xi=re^{i\phi}$ and $\alpha=\vert\alpha\vert e^{i\vartheta}$
parameterize the squeezing and displacement, respectively. The resulting
state exhibits a Gaussian Wigner function with quadrature variances
$e^{-2r}$ (squeezed) and $e^{2r}$ (antisqueezed) along the directions
$x_{\phi}=xcos\frac{\phi}{2}+ysin\frac{\phi}{2}$ ($y_{\phi}=-xsin\frac{\phi}{2}+ycos\frac{\phi}{2}$)
when $r>0$. We note that the squeezing operations are the element
of the $SU(1,1)$ group \citep{PhysRevA.33.4033} and can be implemented
by optical parametric amplifiers(OPAs).

As a first demonstration of our protocol, we consider an initial pointer
state prepared as a squeezed coherent state, $\vert\phi\rangle=\vert\alpha,\xi\rangle$.
The resulting output state, derived from Eq. (\ref{eq:8-1}), takes
the form
\begin{align}
\vert\Psi_{1}\rangle & =\mathcal{N}\left[t_{+}D(s)+t_{-}D(-s)\right]|\alpha,\xi\rangle\nonumber \\
 & =\mathcal{N}S\left(\xi\right)\left[t_{+}e^{2isIm(\alpha^{\ast})}|\alpha^{\prime}+s^{\prime}\rangle+t_{-}|\alpha^{\prime}-s^{\prime}\rangle\right],\label{eq:8}
\end{align}
where $\mathcal{N}$ is the normalization constant with 
\begin{equation}
\mathcal{N}^{-2}=|t_{+}|^{2}+|t_{-}|^{2}+2Re\left(t_{+}^{*}t_{-}e^{-2isIm(\alpha^{\ast})}e^{-2Re(\alpha^{\prime})}e^{-2s^{\prime2}}\right),\label{eq:9}
\end{equation}
and the transformed parameters are $\alpha^{\prime}=\alpha cosh(r)+\alpha^{*}e^{i\phi}sinh(r)$
and $s^{\prime}=s\left[cosh(r)+e^{i\phi}sinh(r)\right]$. This expression
reveals that $\vert\Psi_{1}\rangle$ constitutes a squeezed superposition
of coherent states $|\alpha^{\prime}\pm s^{\prime}\rangle$ with relative
weights determined by the weak value $\langle A\rangle_{w}$. These
weights can be tuned by adjusting the pre- and post-selection parameters
of the measured system, while the structure of the superposition depends
on the interaction strength $s$. Consequently, $\vert\Psi_{1}\rangle$
can be continuously tuned across a family of nG states by varying
these parameters. 

For analytical clarity, we consider the special case of a squeezed
vacuum input ($\vert\alpha\vert=0$) with $\phi=0$, yielding the
simplified output state 

\begin{equation}
\vert\Psi_{2}\rangle=\mathcal{\eta}S\left(r\right)\left[t_{+}|se^{r}\rangle+t_{-}|-se^{r}\rangle\right]\label{eq:10-2}
\end{equation}
with normalization 
\begin{equation}
\eta^{-2}=2\left[\left(1+e^{-2s^{2}e^{2r}}\right)+|\langle A\rangle_{\omega}|^{2}\left(1-e^{-2s^{2}e^{2r}}\right)\right].\label{eq:11-1}
\end{equation}
This form enables the direct observation of the measurement's effect
on the squeezed vacuum. By varying $\langle A\rangle_{w}$ and $s$,
one obtains a continuous family of intermediate squeezed cat-like
states, whose properties we examine next.

\begin{figure}
\includegraphics[width=8cm]{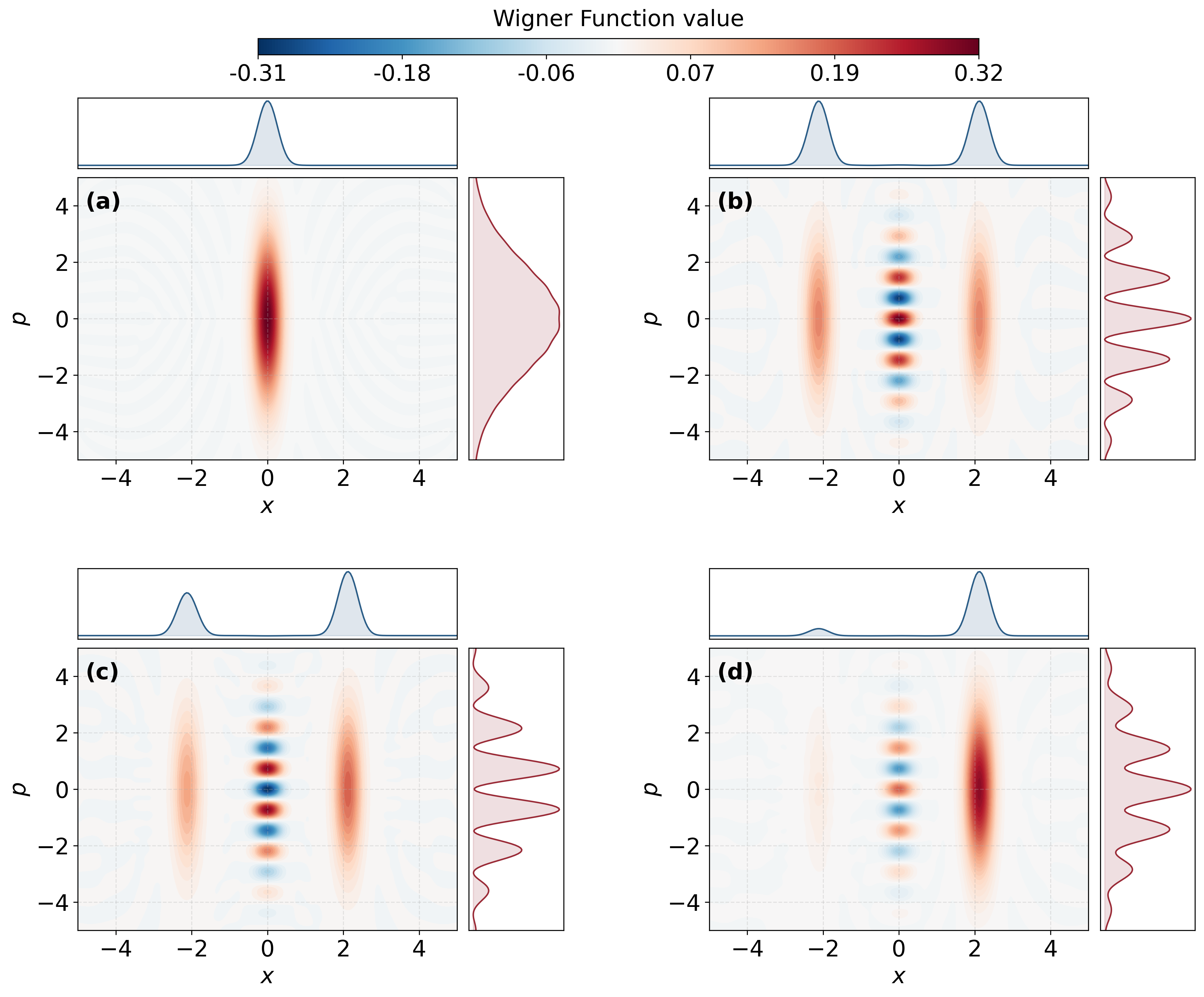}

\caption{\label{fig:2} The Wigner function of the output state $\vert\Psi_{2}\rangle$
for different system parameters. (a) for input state $\vert\xi\rangle$;
(b) for $\langle A\rangle_{w}=0$; (c) for $\langle A\rangle_{w}=10$;
(d) for $\langle A\rangle_{w}=0.5$. Other parameters are taken as
$s=1.5$, $r=1$, and $\phi=0$. }
\end{figure}

To characterize the state $\vert\Psi_{2}\rangle$, we analyze its
Wigner function for various parameter configurations in Fig. \ref{fig:2}.
Unlike the input squeezed vacuum state, the output state exhibits
squeezing along both quadratures. Notably, when $\langle A\rangle_{w}=0$
(see Fig. \ref{fig:2} (b)), the state displays highly squeezed peaks
in both quadrature directions, forming a close approximation to a
GKP state \citep{PhysRevA.64.012310,PhysRevLett.128.170503}. For
other parameter values, the protocol generates intermediate GKP-like
states exhibiting comb structures in both position and momentum quadratures
(see the Fig. \ref{fig:2} (c) and (d)). Given the importance of GKP
states in optical quantum computing, our approach offers an alternative
methodology for implementing these resource states with enhanced efficiency.

As established in previous studies \citep{PhysRevA.78.063811,PhysRevA.111.043704},
the squeezed cat state corresponding to $\langle A\rangle_{w}=0$
in our model can be approximated as $S(r^{\prime})\left[c_{0}\vert0\rangle+c_{2}\vert2\rangle\right]$
with $c_{0},c_{2}\in\mathbb{C}$ for sufficiently large amplitudes.
Algebraically, this state can be obtained through photon subtraction
operations on a squeezed vacuum state:
\begin{align}
a^{2}S(r^{\prime})\vert0\rangle & \propto S(r^{\prime})\left[\vert0\rangle-\sqrt{2}\tanh r^{\prime}\vert2\rangle\right],\label{eq:13}\\
a^{\dagger}aS(r^{\prime})\vert0\rangle & \propto S(r^{\prime})\left[\vert0\rangle-\sqrt{2}\left(\tanh r^{\prime}\right)^{-1}\vert2\rangle\right].
\end{align}
 Thus, our generation method for standard squeezed cat states is formally
equivalent to conventional photon-subtraction techniques. However,
traditional approaches require N-photon subtraction to achieve high-fidelity
( $F>0.999$ ) large-amplitude squeezed cat states, which involves
probabilistic processes with inherently low success rates. These methods
typically employ heralding schemes based on photon detection, where
photon subtraction is implemented via beam splitters whose transmissivity
parameters further diminish the overall success probability. This
efficiency reduction becomes particularly severe for multi-photon
($N\ge2$) subtraction protocols. 

\begin{figure}
\includegraphics[width=8cm]{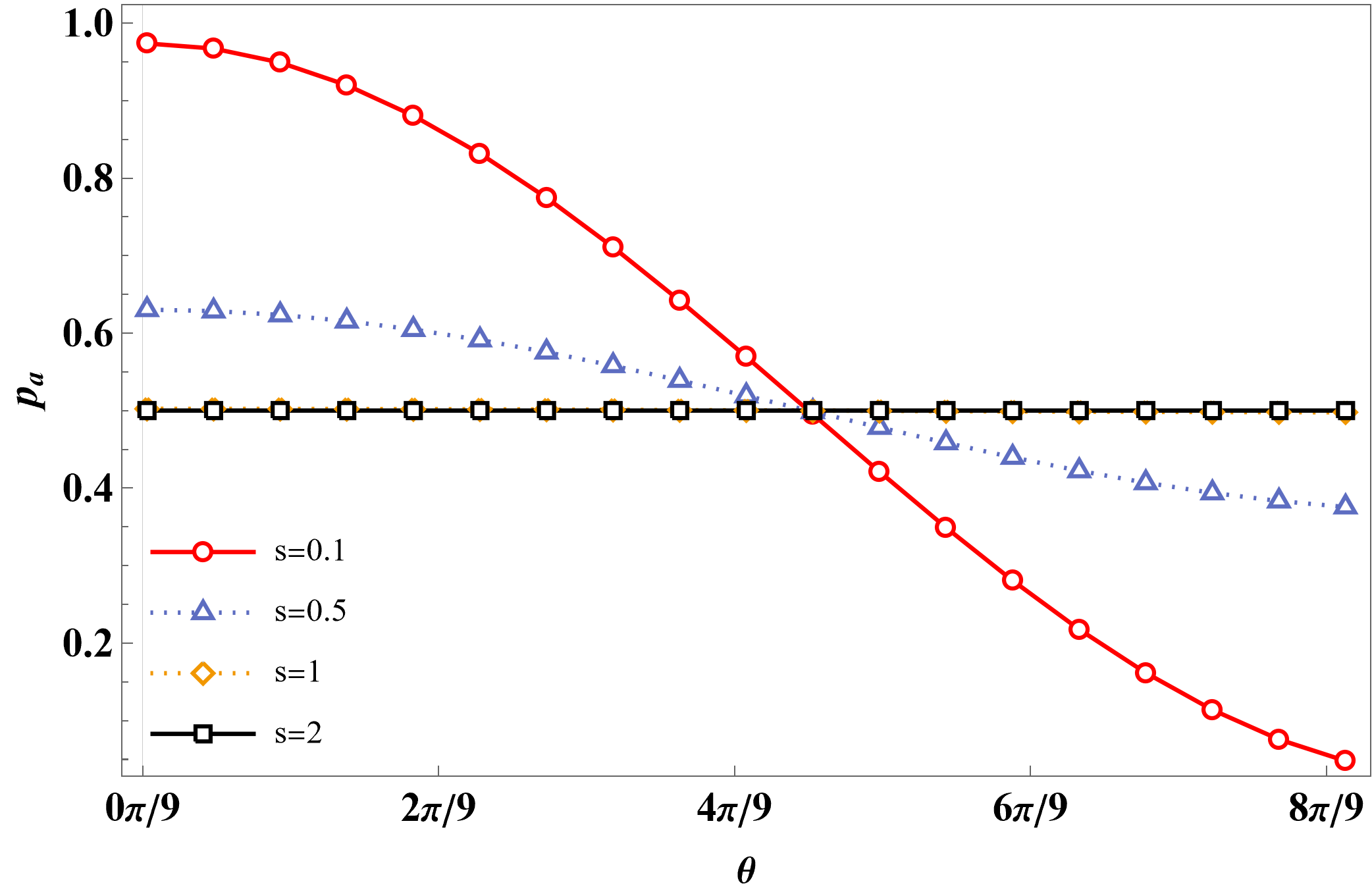}

\caption{\label{fig:3}The success probability of the output state $\vert\Psi_{2}\rangle$
as a function of the weak value parameter $\theta$ for different
interaction strength parameters $s$. Here, we take $r=0.5$ $\delta=0$,
and $\phi=0$. }

\end{figure}

Our protocol effectively addresses the limitations of traditional
methods for generating squeezed cat states. Since $e^{r}$ assumes
strictly positive values, the effective amplitude of the coherent
state superposition $\vert\pm se^{r}\rangle$ can be significantly
enhanced through squeezing ($r>0$), even for moderate interaction
strengths $s$. For example, with parameters $s=1$, $\langle A\rangle_{w}=0$
and $r=1$, our protocol generates a large-amplitude ($\alpha=2.71$)
single-mode squeezed even cat state. 

Furthermore, as shown in Fig. \ref{fig:3}, the successful probability
$p_{a}=\frac{1}{4}\vert\langle\psi_{f}\vert\psi_{i}\rangle\vert^{2}\eta^{-2}$
remains substantial. For this analysis, we assume $A=\sigma_{x}$
and use the parameterization described at the end of Section \ref{sec:2}.
Although large anomalous weak values typically correspond to diminished
postselection probabilities, our von Neumann-type measurement scheme
enables the generation of highly nG states with appreciable success
rates when employing larger values of both the interaction strength
$s$ and squeezing parameter $r$ (see Fig. \ref{fig:3}). 

\subsection{\label{subsec:B}Typical Cat-like states}

Quantum cat states represent a significant class of nonclassical states
in quantum optics, defined as superpositions of coherent states with
opposite phases: $\vert CS_{\pm}\rangle=N\left(\vert\alpha\rangle\pm e^{i\theta}\vert-\alpha\rangle\right)$,
where $N$ denotes the normalization coefficient. The generation and
amplification of such states have been extensively investigated across
various physical platforms \citep{ourjoumtsev2007generation,2023}.
In this subsection, we demonstrate the preparation of generalized
cat states using our postselected von Neumann measurement framework.

By setting the squeezing parameter to zero ($\xi=0$) in the protocol
described in Sec.\ref{subsec:A}, the input pointer state reduces
to a coherent state $\vert\phi\rangle=\vert\alpha\rangle=D(\alpha)\vert0\rangle$.
The corresponding output state becomes
\begin{equation}
\vert\Psi_{3}\rangle=\gamma\left[t_{+}|\alpha+s\rangle+t_{-}|\alpha-s\rangle\right],\label{eq:10-1}
\end{equation}
with normalization constant 
\begin{equation}
\gamma^{-2}=2\left[\left(1+e^{-2s^{2}}\right)+|\langle A\rangle_{w}|^{2}\left(1-e^{-2s^{2}}\right)\right].\label{eq:10}
\end{equation}
This represents a cat-like state comprising a superposition of displaced
coherent states $\vert\alpha\pm s\rangle=D(\alpha\pm s)\vert0\rangle$.
Unlike symmetric cat states, the weights of the two coherent components
in our generated states are tunable and asymmetric. This asymmetry
becomes negligible for small $\vert\alpha\vert$. Since the weak value
$\langle A\rangle_{w}$ can assume arbitrary complex values, continuous
variation of this parameter enables the generation of both standard
cat states and intermediate variants, all exhibiting nG characteristics
except for the special cases $\langle A\rangle_{w}=\pm1$. 

\begin{figure}
\includegraphics[width=8cm]{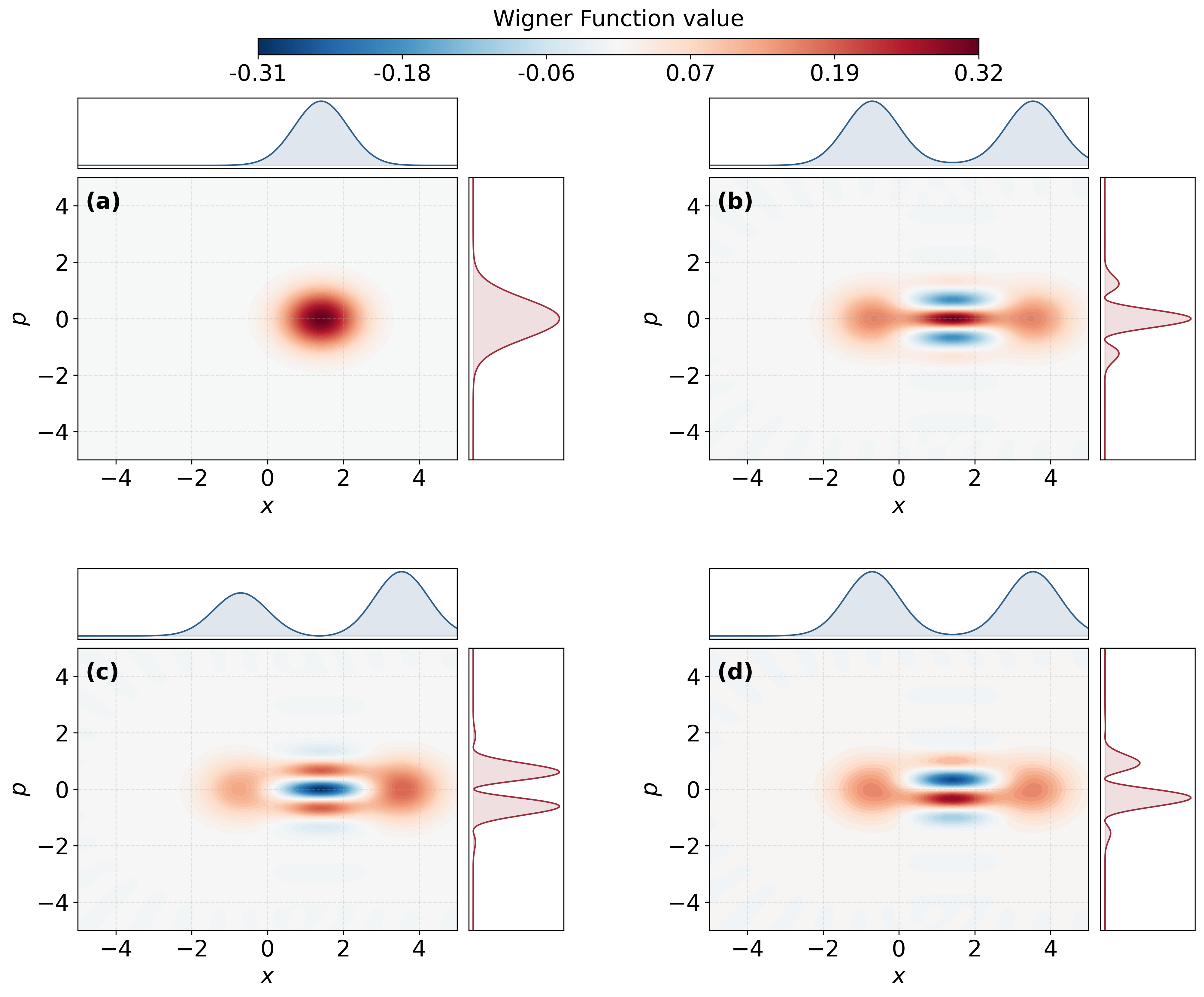}

\caption{\label{fig:3-2}The Wigner function of $\vert\Psi_{3}\rangle$ for
different weak values. (a) for initial input coherent state; (b) for
$\langle A\rangle_{w}=0$; (c) for $\langle A\rangle_{w}=10$ and
(d) for $\langle A\rangle_{w}=-i$. Here, we take $\vert\alpha\vert=1$,
$\vartheta=0$ and $s=1.5$.}
\end{figure}

To elucidate the effects of the postselected von Neumann measurement
on coherent states, Fig. \ref{fig:3-2} presents Wigner function visualizations
of $\vert\Psi_{3}\rangle$ for different weak values with fixed parameters
$\vert\alpha\vert=1$ and $s=1.5$. These plots confirm that our measurement
scheme generates various cat-like states exhibiting characteristic
Wigner function negativities. As predicted, the morphology of quantum
interference fringes in phase space depends systematically on the
weak value of the system observable.

When the input state is a fundamental Gaussian beam corresponding
to $\vert\alpha\vert=0$ (the vacuum state), the pointer state reduces
to

\begin{align}
\vert\Psi_{4}\rangle & =\gamma\left(t_{+}|s\rangle+t_{-}|-s\rangle\right).\label{eq:9-1}
\end{align}
 Remarkably, this demonstrates that even when initializing from the
vacuum state, our protocol enables the preparation of cat-like states.

We now detail the generation of three characteristic cat-state variants
using our protocol: 

(i) For $\langle A\rangle_{w}=0$, Eq. (\ref{eq:9-1}) reduces to
the standard even cat state
\begin{equation}
\vert\phi\rangle_{even}=\frac{1}{\sqrt{2\left(1+e^{-2s^{2}}\right)}}\left(\vert s\rangle+\vert-s\rangle\right).\label{eq:11}
\end{equation}
(ii) When $\langle A\rangle_{w}\gg1$, the protocol generates an (unnormalized)
odd cat-like state of the form $b_{1}\vert s\rangle-b_{2}\vert-s\rangle$,
where $b_{1},b_{2}>0$. 

(iii) Setting $\langle A\rangle_{w}=-i$ yields the Yurke-Stoler state:
\begin{align}
\vert\phi\rangle_{ys} & =\frac{1}{\sqrt{2}}\left(|s\rangle+i|-s\rangle\right).\label{eq:o}
\end{align}
The continuous tunability of the interaction strength parameter $s$
enables the controlled amplitude scaling of the coherent state components,
facilitating observation of the transition from \textquotedbl kitten\textquotedbl{}
states (small amplitude) to \textquotedbl big cat\textquotedbl{}
states (large amplitude). On the other hand, the above enlarged typical
cat states can also be obtained from Eq. (\ref{eq:10-2}) by unsqueezing
the state $\vert\Psi_{2}\rangle$ \citep{h2x6-dz96}. We can achieve
the inverse squeezing operator $S^{-1}(r)$ by using an OPA with its
pump phase shifted by 180 degrees compared to the OPA that implemented
the original squeezing $S(r)$. This is a standard and experimentally
well-established technique in quantum optics. 

\begin{figure}
\includegraphics[width=8cm]{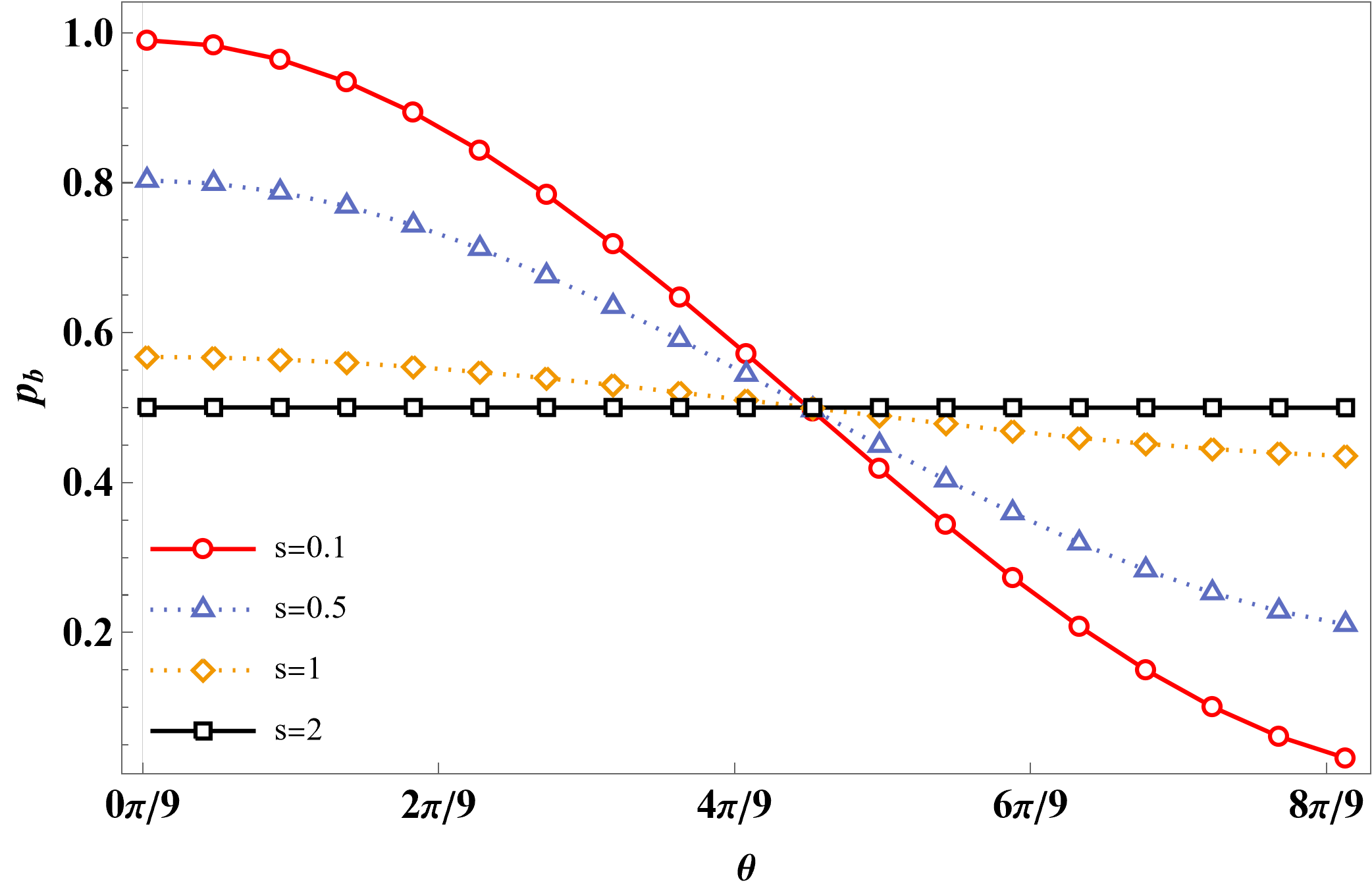}

\caption{\label{fig:T}The success probability of postselection of the final
pointer state $\vert\Psi_{4}\rangle$ as a function of weak value
parameter $\theta$ for different interaction coupling strength $s$.
Here, we take $\delta=0$. }
\end{figure}

Figure \ref{fig:T} illustrates the dependence of the success probability
$p_{b}$ for generating $\vert\Psi_{4}\rangle$ on both the weak value
$\langle\sigma_{x}\rangle_{w}$ (as defined in Section \ref{sec:2})
and interaction strength $s$. In the weak coupling regime ( $0<s<1$),
the success probability decreases with increasing $s$, with the rate
of decrease being proportional to the magnitude of anomalous weak
values. However, in the strong interaction regime ( $s>1$), the probability
becomes largely independent of $s$ since $e^{-2s^{2}}\approx0$ for
$s\gg1$. For instance, at $s=2$, the success probability remains
essentially constant ($p_{b}\approx0.5$ ) across all values of the
weak value parameters $\theta$ and $\delta$. 

Experimental implementation of these single mode cat-like states can
be achieved using optical and trapped-ion systems. For example, recent
work by Pan et al. \citep{Nature2020} demonstrated weak-to-strong
measurement transitions in a trapped-ion platform, whose framework
can be adapted to generate our output state $\vert\Psi_{4}\rangle$.
In their setup, the vibrational motion of a single trapped ion---initially
prepared in a fundamental Gaussian profile---serves as the pointer,
coupled to the ion's internal electronic states via a bichromatic
light field. This configuration realizes the interaction Hamiltonian
$H=g\sigma_{x}\otimes P$. States analogous to our $\vert\Psi_{4}\rangle$
can be generated by appropriate selection of pre- and post-selected
states for the internal electronic degrees of freedom, as exemplified
in Eq. (5) of Ref. \citep{Nature2020}. Consequently, the transition
from kitten to big cat states can be observed by tuning the interaction
strength between the ion's internal state and its vibrational motion
using established experimental techniques. 

In summary, our protocol enables the conversion of initially Gaussian
states into nG states via postselected von Neumann measurements. The
generated cat-like states can be systematically manipulated through
controlled adjustment of the interaction strength $s$ and weak value
$\langle A\rangle_{w}$. 

\section{\label{sec:4}Generation of Two-Mode NG States}

The two-mode Schr�dinger cat state represents an important quantum
resource, initially proposed in Refs. \citep{PhysRevLett.71.2360,2025},
and is defined as either $\vert\psi\rangle=\mathbb{N}(\vert\alpha\rangle\vert\beta\rangle+e^{i\theta}|-\alpha\rangle\vert-\beta\rangle)$
or $\vert\psi\rangle=\mathbb{N}(\vert\alpha\rangle\vert-\beta\rangle+e^{i\theta}\vert-\alpha\rangle\vert\beta\rangle)$,
where $\mathbb{N}$ denotes the normalization constant. Our protocol
can also generate entangled two-mode squeezed Schr�dinger cat states.
Below, we detail the generation of such two-mode entangled cat-like
states using our approach.

\begin{figure}[H]
\centering
\includegraphics[width=8cm]{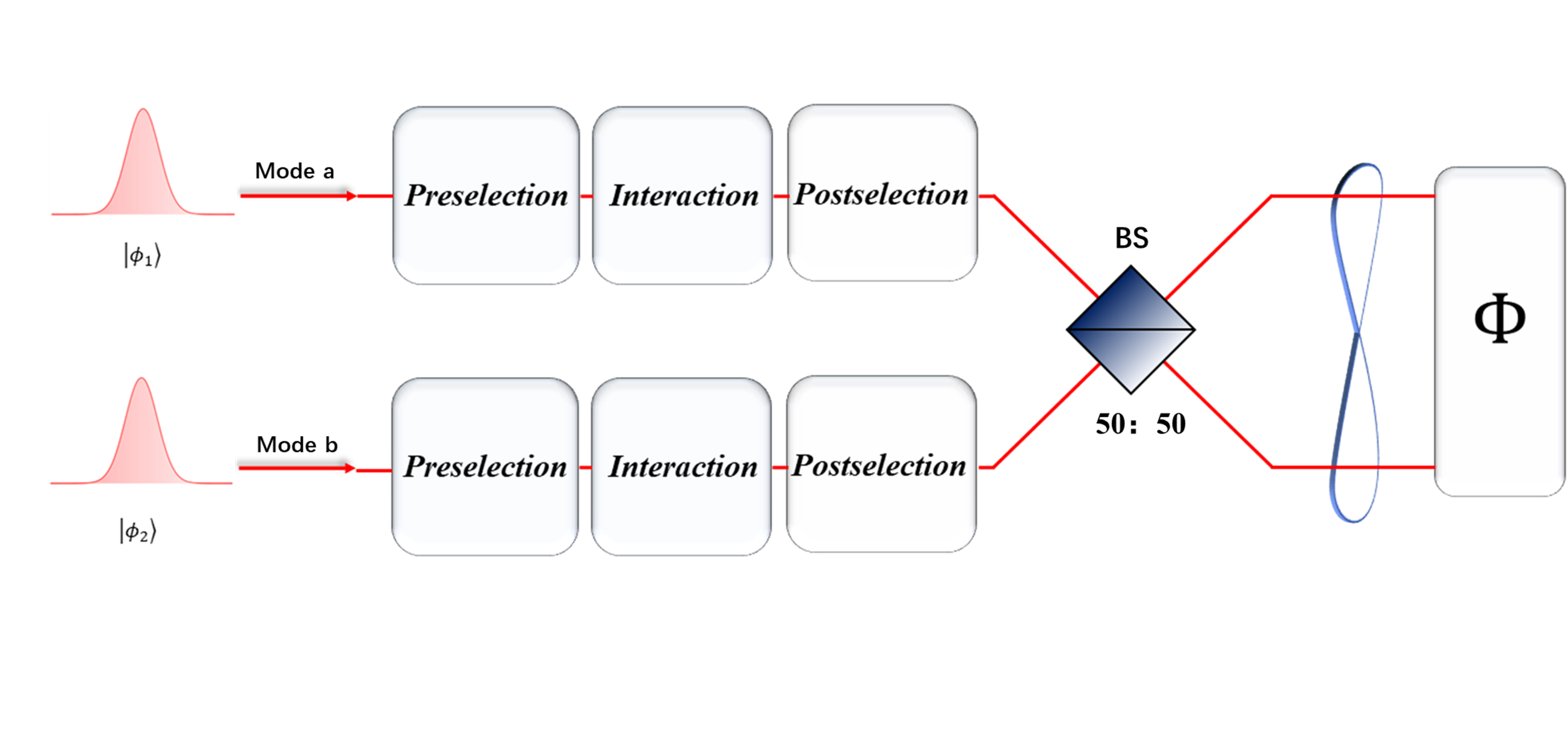}

\caption{\label{fig:5-1}This schematic diagram illustrates the protocol for
preparing two-mode entangled nG states via postselected von Neumann
measurement. The system is initially prepared in two spatially separated
Gaussian states. After undergoing postselected measurement and interaction
with a BS, two-mode entangled nG states can be generated.}
\end{figure}

The schematic of our proposal is presented in Fig. \ref{fig:5-1}.
As shown, two prepared single-mode states (which may be identical
or different) are fed into our protocol, with one or both modes undergoing
postselected von Neumann measurements. After passing through a 50:50
beam splitter (BS), a two-mode entangled nG state is generated with
a definite probability. The postselected von Neumann measurement procedure
is identical to the single-mode case described in Sec. \ref{sec:2}. 

A crucial element of our protocol is the BS, which enables the entanglement
generation between the two subsystems. The BS operation is described
by the scattering matrix \citep{PhysRevA.40.1371}
\begin{equation}
U_{BS}=\left(\begin{array}{cc}
\cos\tau e^{i\varphi_{\tau}}\ \  & \sin\tau e^{i\phi_{\rho}}\ \\
-\sin\tau e^{-i\phi_{\rho}} & \cos\tau e^{-i\varphi_{\tau}}
\end{array}\right).\label{eq:16}
\end{equation}
The reflectivity and transmittance are defined as $\sqrt{R}=\sin\tau e^{i\phi_{\rho}}$
and $\sqrt{T}=\cos\tau e^{i\varphi_{\tau}}$, respectively, satisfying
$R+T=1$. For a balanced $50:50$ BS, we select parameters $\varphi_{\tau}=\phi_{\rho}=0$,
$\tau=\frac{\pi}{4}$.

\subsection{\label{subsec:A-1}Two-mode entangled squeezed cat-like state}

As a specific example, we consider an input pointer state comprising
two squeezed vacuum (SV) states, $|\phi\rangle=\vert\phi_{1}\rangle\vert\phi_{2}\rangle=|\xi\rangle_{a}|-\xi\rangle_{b}$.
We then apply a postselected von Neumann measurement to only one mode
(the $a$-mode) with the interaction Hamiltonian $H_{a}=gA\otimes P_{a}$.
We also assume that the internal degree of freedom of the beam corresponding
to the $a$-mode is prepared in the state $\vert\psi_{i}\rangle$.
Following time evolution and postselection onto the final system state
$|\psi_{f}\rangle$, the initial composite state transforms as 
\begin{align}
\vert\varPhi_{evol}\rangle & =\langle\psi_{f}\vert U_{a}|\psi_{i}\rangle\otimes|\xi\rangle_{a}\vert-\xi\rangle_{b}\nonumber \\
 & =\frac{1}{2}\langle\psi_{f}|\psi_{i}\rangle S_{a}(\xi)\left[t_{+}|s^{\prime}\rangle_{a}+t_{-}|-s^{\prime}\rangle_{a}\right]|-\xi\rangle_{b}.\label{eq:13-1}
\end{align}
 At this stage, mode $b$ remains unmodified. To establish correlation
between the modes, the state is passed through a balanced $50:50$
BS, yielding:

\begin{align}
\vert\varPhi_{1}\rangle & =\mathcal{\eta}S_{ab}(\xi)\left[t_{+}\left|\frac{s^{\prime}}{\sqrt{2}}\right\rangle _{a}\left|-\frac{s^{\prime}}{\sqrt{2}}\right\rangle _{b}+t_{-}\left|-\frac{s^{\prime}}{\sqrt{2}}\right\rangle _{a}\left|\frac{s^{\prime}}{\sqrt{2}}\right\rangle _{b}\right],\label{eq:19}
\end{align}
where the normalization constant $\mathcal{\eta}$ is identical to
that in Eq. (\ref{eq:11-1}). Here, $S_{ab}(\xi)=exp(\xi^{*}ab-\xi a^{\dag}b^{\dag})$
denotes the two-mode squeezing operator, which simultaneously squeezes
and entangles the modes. The resulting state $\vert\varPhi_{1}\rangle$
represents a family of two-mode entangled nG cat-like states for all
$\langle A\rangle_{w}\neq\pm1$.

Alternatively, applying the postselected measurement to mode $b$
with input state $|-\xi\rangle_{a}|\xi\rangle_{b}$ produces
\begin{equation}
\vert\varPhi_{2}\rangle=\mathcal{\eta}S_{ab}(\xi)\left[t_{+}\left|\frac{s^{\prime}}{\sqrt{2}}\right\rangle _{a}\left|\frac{s^{\prime}}{\sqrt{2}}\right\rangle _{b}+t_{-}\left|-\frac{s^{\prime}}{\sqrt{2}}\right\rangle _{a}\left|-\frac{s^{\prime}}{\sqrt{2}}\right\rangle _{b}\right].\label{eq:18-1}
\end{equation}
Thus, our protocol converts uncorrelated squeezed vacuum states into
two-mode entangled cat-like states, with relative weights and amplitudes
controlled by $\langle A\rangle_{w}$ and $s$, respectively. 

The states $\vert\Phi_{1}\rangle$ and $\vert\Phi_{2}\rangle$ encompass
both standard two-mode entangled cat states and intermediate variants,
with components proportional to $1\pm\langle A\rangle_{w}$. By appropriately
adjusting the weak value enables the preparation of diverse entangled
cat states. Specifically, for $\langle A\rangle_{w}=0$ and $\phi=0$,
these states reduce to
\begin{equation}
\vert\varPhi_{1}^{\prime}\rangle=\kappa S_{ab}(r)\left[\left|\frac{se^{r}}{\sqrt{2}}\right\rangle _{a}\left|-\frac{se^{r}}{\sqrt{2}}\right\rangle _{b}+\left|-\frac{se^{r}}{\sqrt{2}}\right\rangle _{a}\left|\frac{se^{r}}{\sqrt{2}}\right\rangle _{b}\right],\label{eq:24-1}
\end{equation}
and 
\begin{equation}
\vert\varPhi_{2}^{\prime}\rangle=\kappa S_{ab}(r)\left[\left|\frac{se^{r}}{\sqrt{2}}\right\rangle _{a}\left|\frac{se^{r}}{\sqrt{2}}\right\rangle _{b}+\left|-\frac{se^{r}}{\sqrt{2}}\right\rangle _{a}\left|-\frac{se^{r}}{\sqrt{2}}\right\rangle _{b}\right]\label{eq:25-1}
\end{equation}
with $\kappa=1/\sqrt{2(1+e^{-2s^{2}e^{2r}})}$. These represent nG
squeezed identical twin cat states, whose amplitudes scale exponentially
with $r>0$. For $s^{2}e^{2r}\ge4$, these states approximate identical
cat states distributed across two modes \citep{wang2016schrodinger}.
Notably, the unitarity of BS preserves generation probabilities, so
the success rates for $\vert\varPhi_{1}\rangle$ and $\vert\varPhi_{2}\rangle$
match those of the single-mode case {[}see the Fig. \ref{fig:3}{]}. 

To quantify non-Gaussianity, we analyze the joint Wigner function
of $\vert\varPhi_{2}\rangle$ which is defined as

\begin{align}
W_{J}\left(\alpha,\beta\right) & =\frac{4}{\pi^{2}}Tr\left[\rho D(\alpha)D(\beta)P_{j}D^{\dagger}(\beta)D^{\dagger}(\alpha)\right]\nonumber \\
 & =\frac{4}{\pi^{2}}P_{J}(\alpha,\beta),\label{eq:joint wigner functuon}
\end{align}
where $\rho=\vert\Phi_{2}\rangle\langle\Phi_{2}\vert$, $P_{j}=P_{a}P_{b}=e^{i\pi a^{\dagger}a}e^{i\pi b^{\dagger}b}$
is the joint parity operator and $D(\alpha)$ and $D(\beta)$. The
function $P_{J}(\alpha,\beta)$ is bounded by $-1\leq P_{J}(\alpha,\beta)\leq1.$ 

\begin{figure}
\includegraphics[width=8cm]{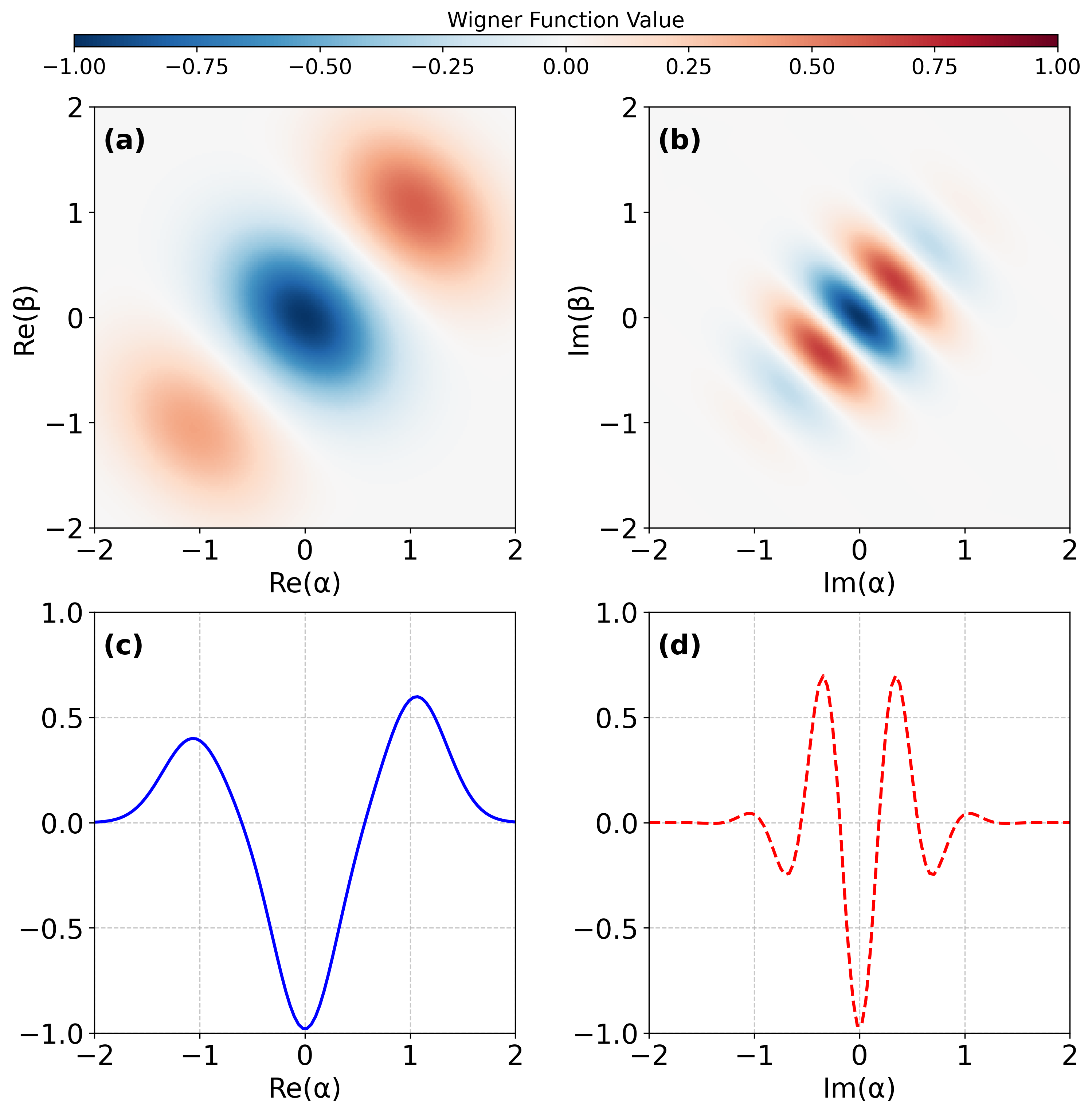}

\includegraphics[width=8cm]{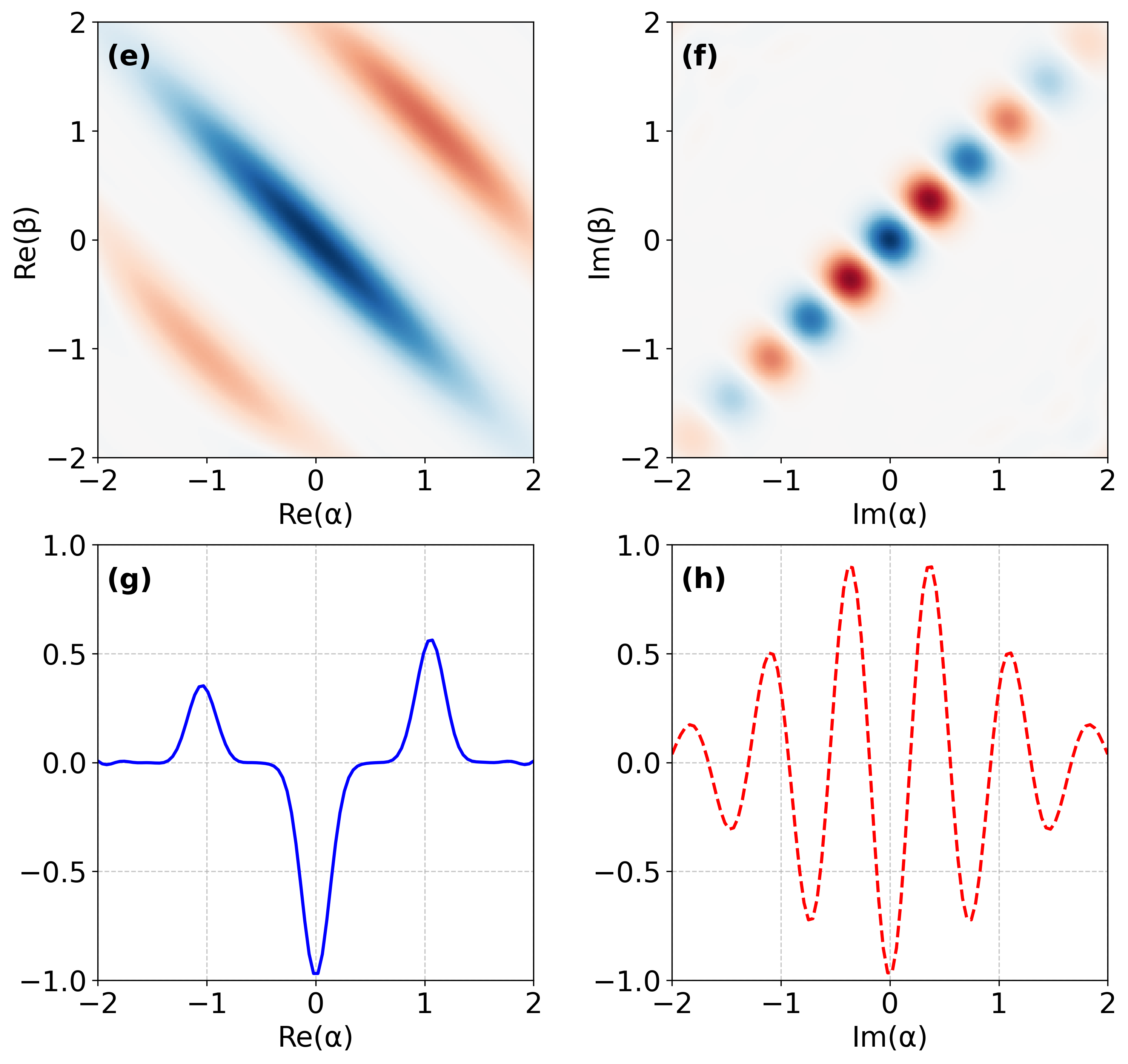}

\caption{\label{fig:6}The scaled joint Wigner function $P_{J}(\alpha,\beta)$
of $\vert\Phi_{2}\rangle$. (a), (b), (e) and (f) represent two-dimensional
projections of the 4D Wigner function for two-mode squeezed odd cat-like
states. Herein, the Panel (a) and (e) ( (b) and (f) ), display the
Wigner function projection with the real parts (imaginary parts) of
$\alpha$ and $\beta$ constrained to $\,Im\left(\alpha\right)=Im\left(\beta\right)=0\,$
($\,Re\left(\alpha\right)=Re\left(\beta\right)=0\,$) for $r=0.2$
and $1.0$, respectively, retaining only their real (imaginary) parts
as variables. (c) and (g) Diagonal line-cuts of the data shown in
(a) and (e), corresponding to 1D plots of the joint Wigner function
along $Re\left(\alpha\right)=Re\left(\beta\right)$ within $Im\left(\alpha\right)=Im\left(\beta\right)=0$.
(d) and (h) Diagonal line-cuts of the data shown in (b) and (f), corresponding
to 1D plots of the joint Wigner function along $Im\left(\alpha\right)=Im\left(\beta\right)$
within $Re\left(\alpha\right)=Re\left(\beta\right)=0$.}

\end{figure}

Figure 7 presents 2D cross-sections of $P_{J}(\alpha,\beta)$ for
different squeezing parameters $r$ with fixed $s=1.5$ and $\langle A\rangle_{\omega}=10$,
respectively. The $\text{Re}(\alpha)$-$\text{Re}(\beta)$ projections
in Fig. \ref{fig:6} (a) and (e) showed central negativity regions
flanked by positive lobes, characteristic of squeezed odd cat-like
states for different squeezing parameters $r$. The projections $\text{Im}(\alpha)$-$\text{Im}(\beta)$
(Fig. \ref{fig:6} (b) and (f) ) exhibit interference fringes with
periodic oscillations, revealing coherent superpositions along the
imaginary axis. Diagonal cuts shown in Fig. \ref{fig:6} (c), (d),
(g) and (h), further highlight these quantum interference effects.
Increasing $r$ enhances the fringe density due to exponential growth
of the effective amplitude $se^{r}$, demonstrating controllable squeezing
and non-Gaussianity.

\subsection{Bell-like state}

Bell states, which originated from the Einstein-Podolsky-Rosen (EPR)
paradox \citep{PhysRev.47.777}, have long served as fundamental examples
for testing quantum theory. As maximally entangled states, they play
crucial roles in quantum key distribution \citep{PhysRevA.65.052331}
and quantum teleportation \citep{WEN2007278}. In continuous-variable
(CV) systems, Bell states can be approximated in the Fock state basis
as \citep{Kanari-Naish_2022}:

\begin{align}
\vert\Psi\rangle_{10} & \propto\text{\ensuremath{\left(\vert\alpha\rangle\vert0\rangle+e^{i\omega}|0\rangle\vert\alpha\rangle\right)}}\label{eq:22-2}\\
\vert\Psi\rangle_{01} & \propto\text{\ensuremath{\left(\vert\alpha\rangle\vert\alpha\rangle+e^{i\omega}|0\rangle\vert0\rangle\right)}}\label{eq:22-1}
\end{align}
Noting that $\langle\alpha|0\rangle=e^{-\frac{1}{2}\vert\alpha\vert^{2}}$,
these states approach maximal entanglement in the coherent-state basis
for large $\vert\alpha\vert$, where the components become nearly
orthogonal. For small $\vert\alpha\vert$ and $\omega=\pi$, they
approximate the number-state Bell states $\vert\Psi\rangle_{\pm}=\frac{1}{\sqrt{2}}\left(\vert0\rangle\vert1\rangle\pm\vert1\rangle\vert0\rangle\right)$.
Here, we demonstrate that our protocol can generate such approximate
Bell states.

Using the setup in Fig. \ref{fig:5-1} with input states $\vert0\rangle_{a}\vert\beta\rangle_{b}$
and postselected measurement on mode $a$, the output state becomes
\begin{equation}
\vert\varPhi_{3}\rangle=\gamma\left[t_{+}|\sqrt{2}s\rangle_{a}|0\rangle_{b}+t_{-}|0\rangle_{a}|\sqrt{2}s\rangle_{b}\right],\label{eq:24}
\end{equation}
where $\gamma$ is the normalization constant and it is consistent
with the expression given in Eq. (\ref{eq:10}). For real $\beta=s$
and large $s$ ($\langle\sqrt{2}s\vert0\rangle=e^{-s^{2}}\thickapprox0$),
this represents a Bell-like state in the coherent-state basis. Mapping
$|\sqrt{2}s\rangle_{a}=\vert1\rangle_{c_{1}}$ and $|0\rangle_{2}=\vert0\rangle_{c_{2}}$
yields: 
\begin{equation}
\vert\varPhi_{3}^{\prime}\rangle=\gamma\left[t_{+}|1\rangle_{c_{1}}|0\rangle_{c_{2}}+t_{-}|0\rangle_{c_{1}}|1\rangle_{c_{2}}\right].\label{eq:28}
\end{equation}
For $\langle A\rangle_{w}=0$ and $\langle A\rangle_{w}\gg1$, this
approximates the Bell states
\begin{equation}
|\psi_{1}\rangle=\frac{1}{\sqrt{2}}\left(|1\rangle_{c_{1}}|0\rangle_{c_{2}}+|0\rangle_{c_{1}}|1\rangle_{c_{2}}\right)\label{eq:33}
\end{equation}
and 
\begin{equation}
|\psi_{2}\rangle\approx\frac{1}{\sqrt{2}}\left(|1\rangle_{c_{1}}|0\rangle_{c_{2}}-|0\rangle_{c_{1}}|1\rangle_{c_{2}}\right),\label{eq:34}
\end{equation}
respectively. 

Similarly, with input $\vert\beta\rangle_{a}\vert0\rangle_{b}$ and
measurement on mode $b$, we obtain:

\begin{equation}
\vert\varPhi_{4}\rangle=\gamma\left[t_{+}|\sqrt{2}s\rangle_{a}|\sqrt{2}s\rangle_{b}+t_{-}|0\rangle_{a}|0\rangle_{b}\right],\label{eq:26}
\end{equation}
which, for $\langle A\rangle_{w}=0$ and $\langle A\rangle_{w}\gg1$
approximates
\begin{equation}
|\psi_{3}\rangle=\frac{1}{\sqrt{2}}\left(|1\rangle_{c_{1}}|1\rangle_{c_{2}}+|0\rangle_{c_{1}}|0\rangle_{c_{2}}\right)\label{eq:38}
\end{equation}
and
\begin{equation}
|\psi_{4}\rangle\approx\frac{1}{\sqrt{2}}\left(|1\rangle_{c_{1}}|1\rangle_{c_{2}}-|0\rangle_{c_{1}}|0\rangle_{c_{2}}\right),\label{eq:38-1}
\end{equation}
respectively. The success probabilities for $\vert\Phi_{3}\rangle$
and $\vert\Phi_{4}\rangle$ remain practical, as shown in Fig. \ref{fig:T}
for specific parameters.

We note that our states $\vert\Phi_{3}\rangle$ and $\vert\Phi_{4}\rangle$
differ from Eqs. (\ref{eq:22-2}) and (\ref{eq:22-1}) by a relative
phase. For small $s$, the vacuum component dominates, preventing
the formation of the number-state basis maximal entanglement Bell
states $\vert\Psi\rangle_{\pm}$. In this work we have focused on
uncorrelated inputs. However, as recent studies showed \citep{PhysRevA.110.052611,tnvf-nrq4,ren2025},
our protocol can also be used to enhance the nonclassicality and non-Gaussianity
of given inherently correlated/entangled two-mode states. 

\subsection{Quantifying Entanglement}

Entanglement, as a fundamental quantum correlation, plays a pivotal
role in quantum information processing and computation, enabling applications
such as quantum teleportation and quantum computing. The quantum states
generated by our protocol contain multiple adjustable parameters and
yield known nG states only under specific parameter configurations.
Consequently, their properties across most parameter regimes remain
largely unexplored. A quantitative characterization of entanglement
is therefore essential for understanding the fundamental features
of these states.

In this subsection, we quantify the entanglement degrees of the generated
two-mode entangled states. Various entanglement measures exist for
two-mode and multi-mode systems \citep{PhysRevLett.78.2275,PhysRevA.57.1619}.
For CV states, concurrence \citep{PhysRevLett.80.2245} and linear
entropy \citep{PhysRevA.75.032301} provide convenient metrics for
quantifying entanglement between subsystems.

We employ linear entropy \citep{PhysRevA.75.032301} to characterize
the entanglement of the two-mode squeezed cat-like states analyzed
in Section \ref{subsec:A-1}. The linear entropy for a bipartite system
is defined as:

\begin{equation}
E=\text{\ensuremath{\frac{d}{d-1}}}\left[1-Tr(\rho_{1}^{2})\right],\label{eq:30}
\end{equation}
where $\rho_{1}$ represents the reduced density matrix of one subsystem,
$Tr(\rho_{1}^{2})$ denotes its purity, and $d=dim(\rho_{1})$ is
the dimension of the subsystem's Hilbert space. For CV states, the
linear entropy satisfies $0\leqslant E\leqslant1$, where $E=0$ corresponds
to completely separable subsystems and $E=1$ indicates maximal entanglement
(equivalent to a Bell state).

\begin{figure}
\includegraphics[width=4cm]{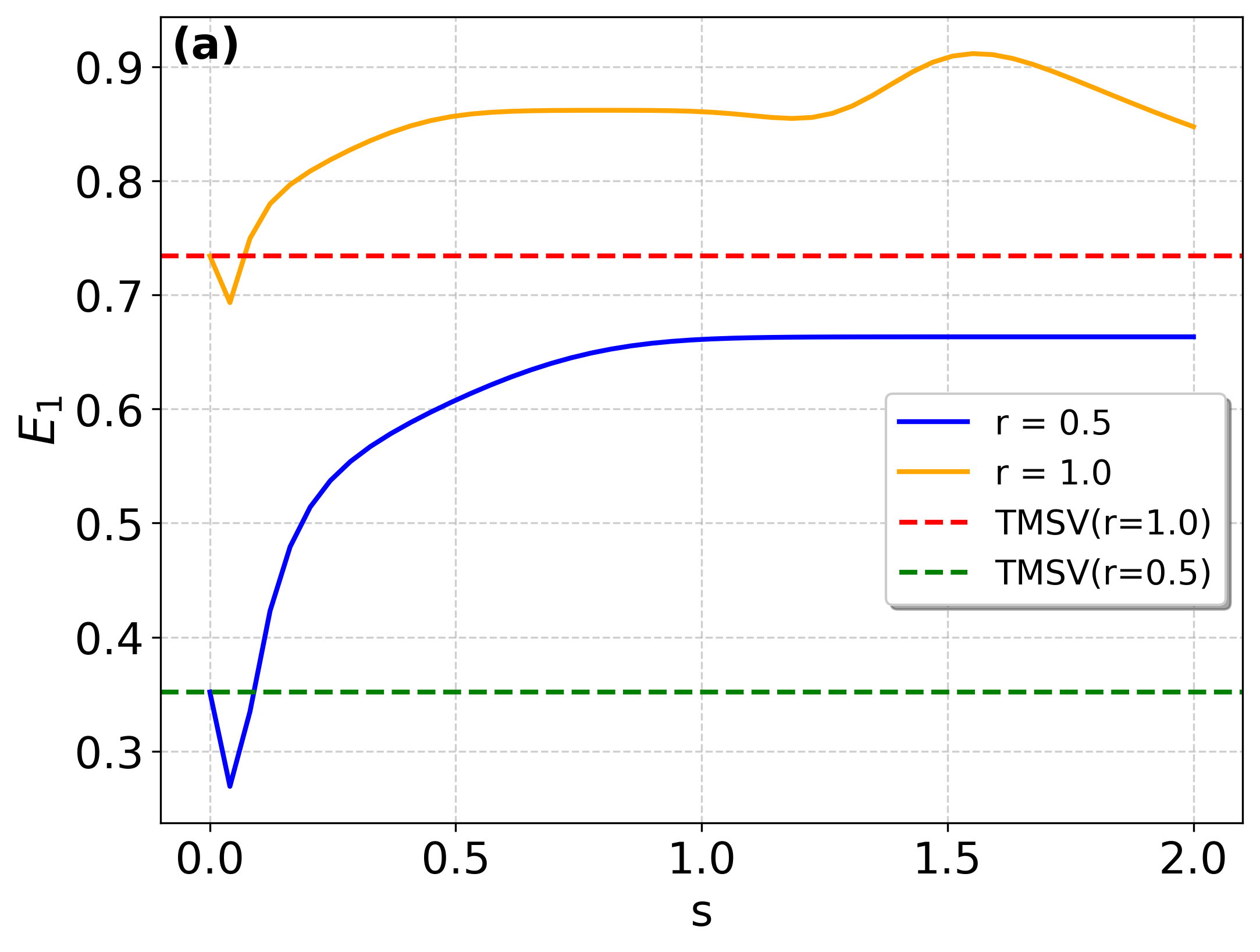}\includegraphics[width=4cm]{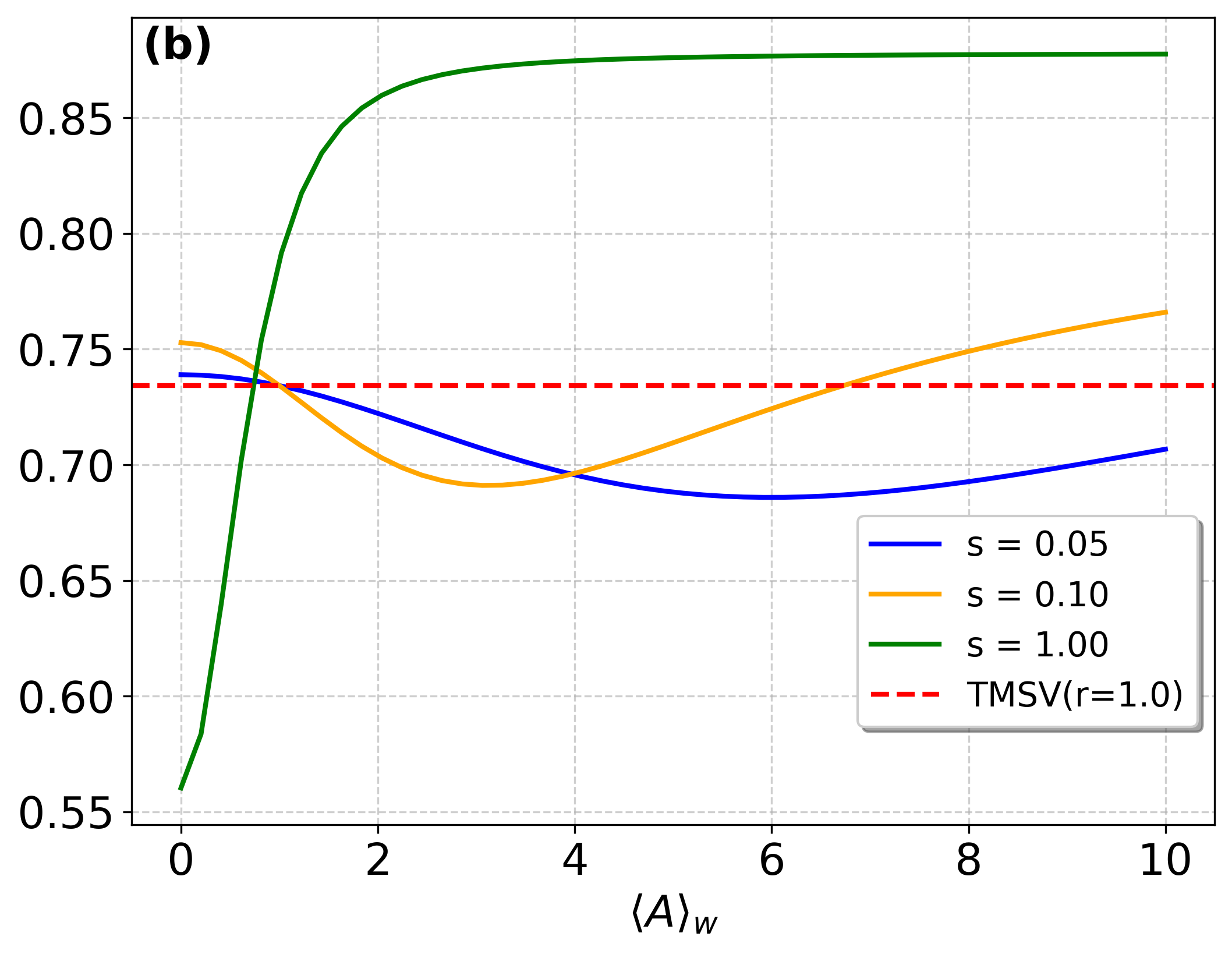}

\includegraphics[width=4cm]{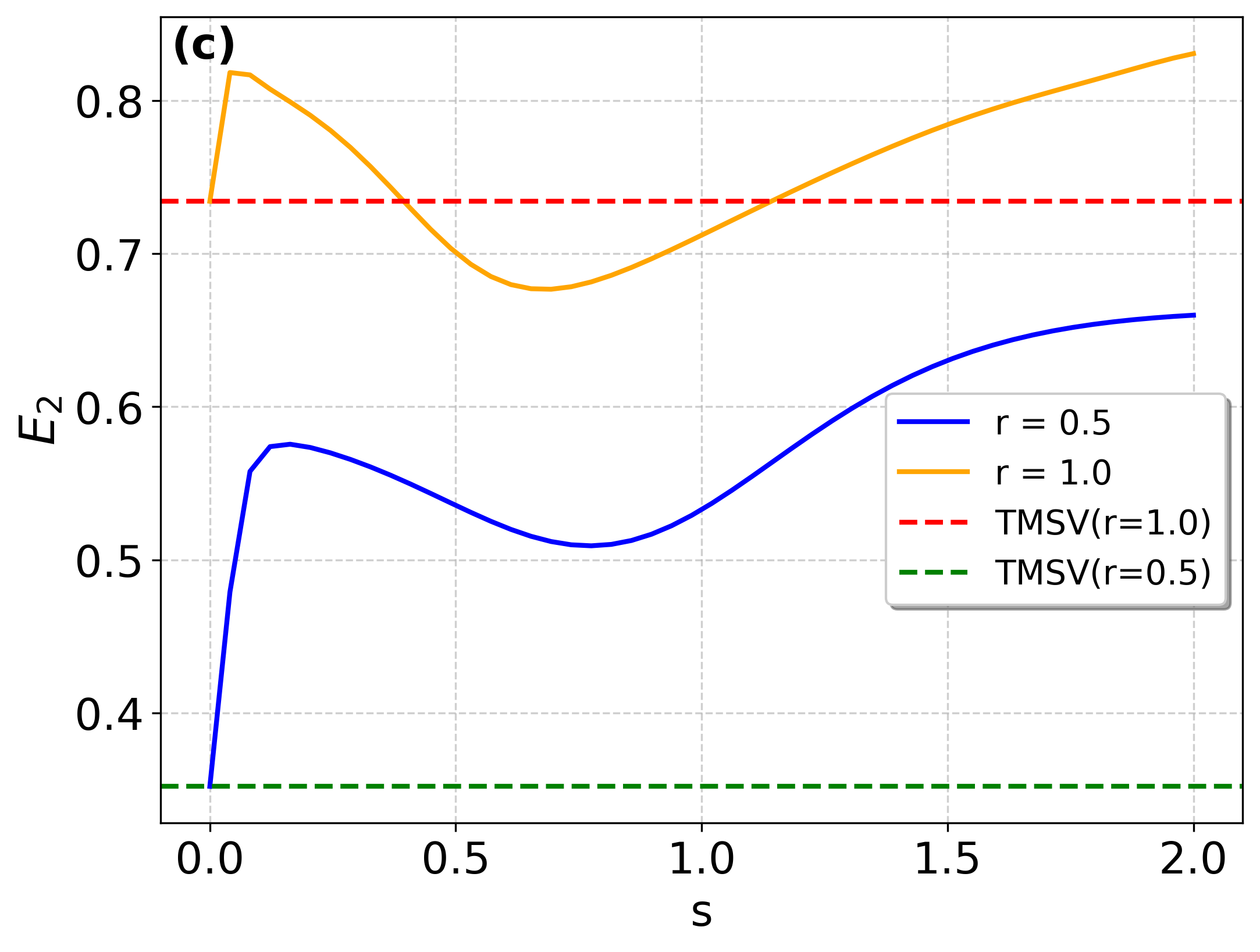}\includegraphics[width=4cm]{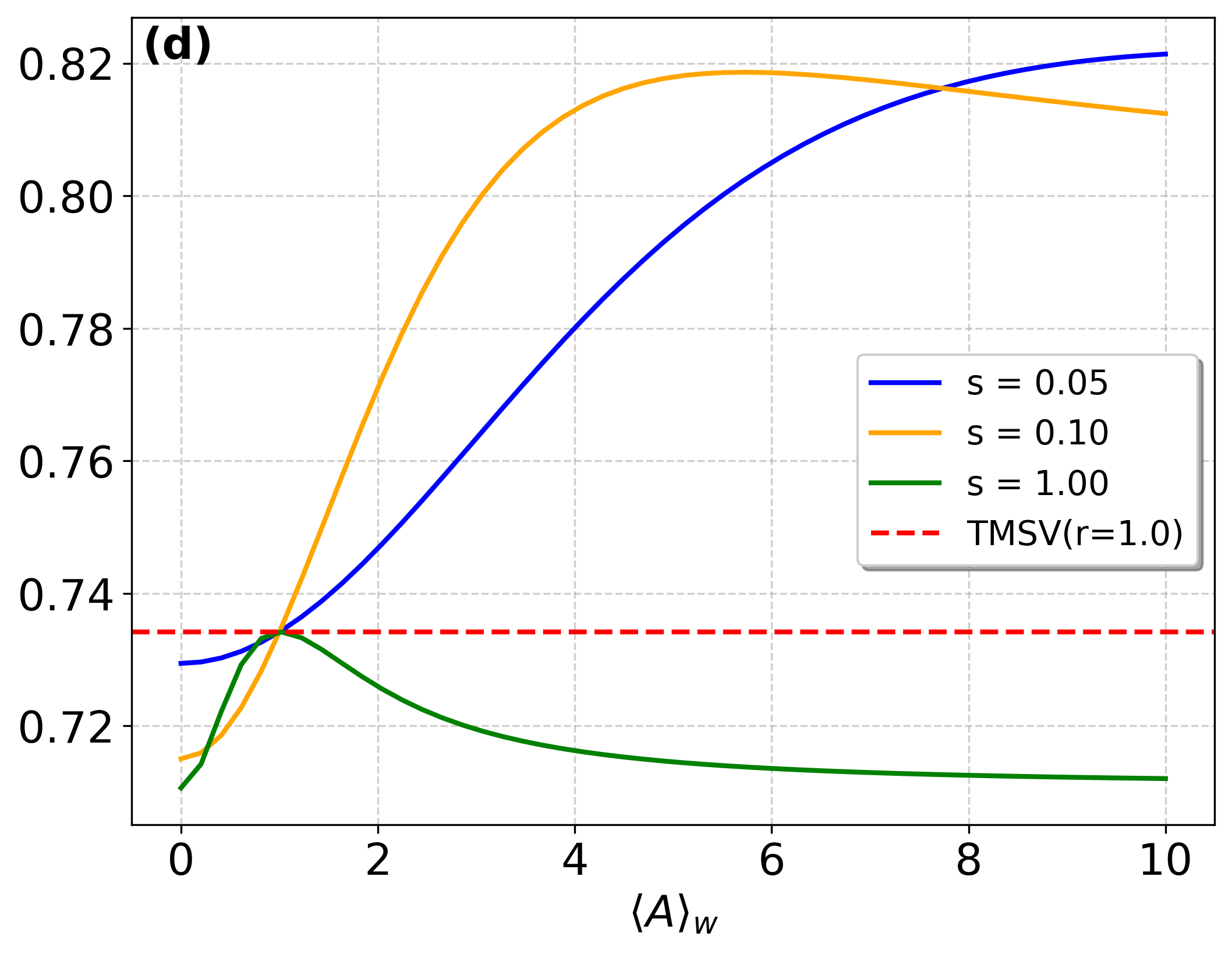}

\caption{\label{fig:8}(a) \& (b) and (c) \& (d) depict the linear entropy
of $\vert\Phi_{1}\rangle$ and $\vert\Phi_{2}\rangle$, respectively.
For (a) and (c), $\langle A\rangle_{\omega}$ is fixed at $10$, and
plot the curves of linear entropy varying with $s$ under the conditions
of $r=0.5$ and $r=1.0$, respectively. In (b) and (d), with $r=1.0$,
the curves of linear entropy varying with the weak value are plotted
under different coupling strengths ( $s=0.05;0.1;1.0$). The dashed
lines represent the linear entropy of the TMSV state for different
squeezing parameters $r$.}
\end{figure}

Figure \ref{fig:8} presents the entanglement characteristics of states
$\vert\Phi_{1}\rangle$ and $\vert\Phi_{2}\rangle$, quantified through
linear entropy analysis under varying system parameters. For comparative
analysis, we include the linear entropy of the two-mode squeezed vacuum
(TMSV) state, which corresponds to the case without postselected von
Neumann measurement ($s=0$). Figure \ref{fig:8} (a) displays the
linear entropy of $\vert\Phi_{1}\rangle$ as a function of interaction
strength $s$ for different squeezing parameters $r$, with the weak
value fixed at $\langle A\rangle_{w}=10$. The entanglement degree
of $\vert\Phi_{1}\rangle$ exhibits a monotonic increase with $s$,
with this enhancement being particularly pronounced for $r<1$. For
anomalous weak values, the entanglement of $\vert\Phi_{1}\rangle$
surpasses that of the TMSV across most parameter regimes, except in
the limit of extremely small $s$. Figure \ref{fig:8} (b) illustrates
the dependence of entanglement of $\vert\Phi_{1}\rangle$ on weak
value for different $s$ values. In the weak measurement regime ($s\ll1$),
the linear entropy $E_{1}$ remains below that of the TMSV even for
large anomalous weak values, whereas strong interactions ($s>1$)
enable significant entanglement enhancement through large weak values.
Notably, when $\langle A\rangle_{w}=1$ or $-1$ (not shown), the
entanglement degree of $\vert\Phi_{1}\rangle$ equals that of the
TMSV for any fixed $r$, independent of $s$. Figures \ref{fig:8}
(c) and (d) present the corresponding analysis for $\vert\Phi_{2}\rangle$.
The entanglement characteristics of $\vert\Phi_{2}\rangle$ differ
significantly from those of $\vert\Phi_{1}\rangle$: (i) Unlike $\vert\Phi_{1}\rangle$,
the linear entropy $E_{2}$ exhibits non-monotonic behavior with increasing
$s$ in moderate postselection regimes; (ii) Contrary to $\vert\Phi_{1}\rangle$,
the entanglement degree of $\vert\Phi_{2}\rangle$ decreases with
increasing $s$ for large weak values. Based on this comprehensive
numerical analysis, we conclude that $\vert\Phi_{1}\rangle$ exhibits
stronger quantum correlations between modes compared to $\vert\Phi_{2}\rangle$.

\begin{figure}
\includegraphics[width=8cm]{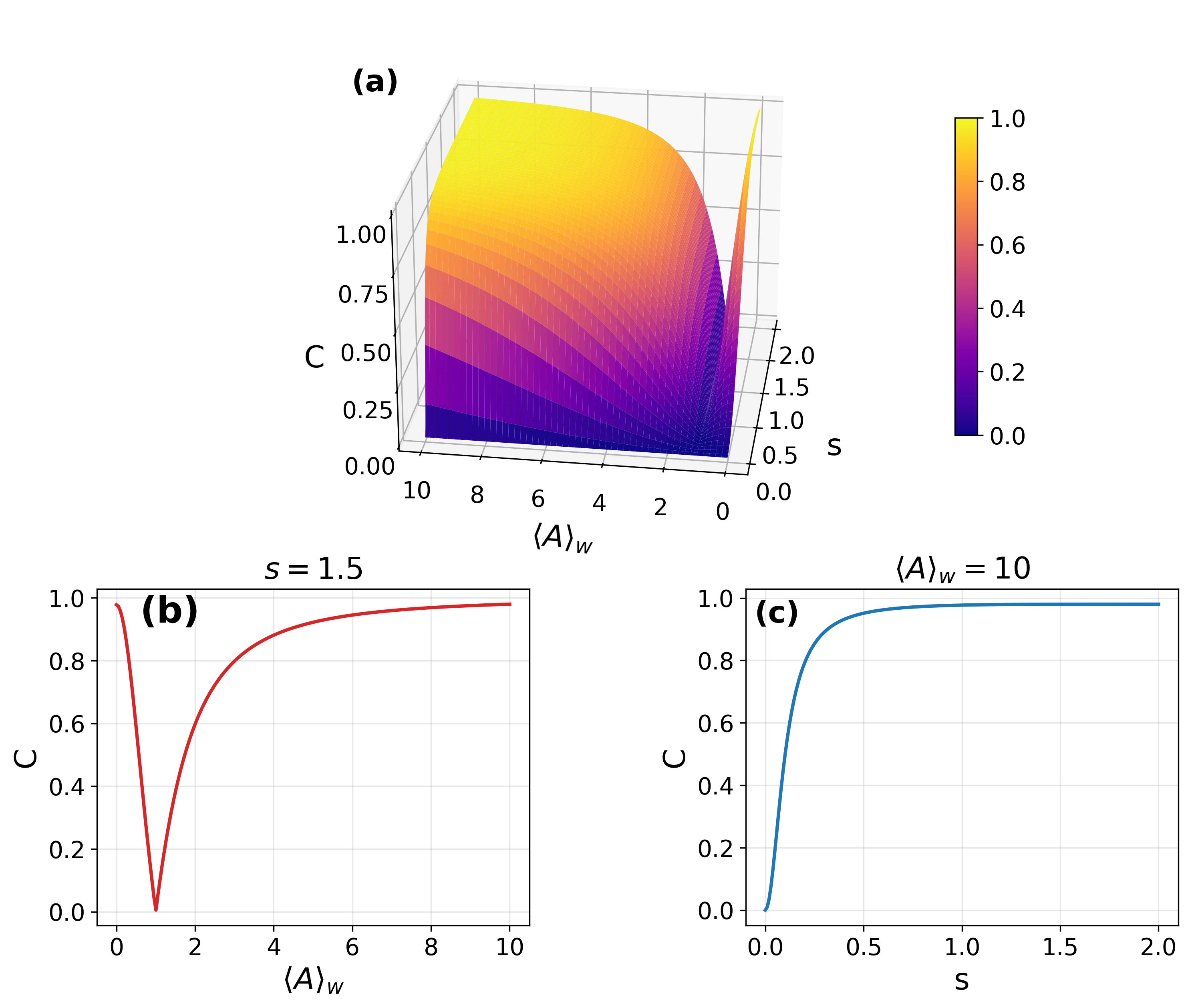}

\caption{\label{fig:7}The concurrence function $C$ for our generated Bell-like
states $\vert\Phi_{3}\rangle$ and $\vert\Phi_{4}\rangle$. Specifically,
(a) is the 3D plot of $C$ as a function of interaction strength parameter
$s$ and the weak value $\langle A\rangle_{w}$ ; the (b) shows the
changes of the concurrence function $C$ with the weak value $\langle A\rangle_{w}$
when the $s$ is fixed to $1.5$; and (c) Concurrence $C$ as a function
of $s$ when the weak value $\langle A\rangle_{w}$ is fixed to 10.}
\end{figure}

For the Bell-like states, we employ concurrence as our entanglement
measure. Concurrence provides a well-established metric for quantifying
entanglement in bipartite systems. Consider a general CV entangled
state of the form:

\begin{equation}
\vert\Psi\rangle=M\left[\lambda\vert a\rangle\varotimes\vert b\rangle+\delta\vert c\rangle\varotimes\vert d\rangle\right],\label{eq:22-3}
\end{equation}
where $\lambda$ and $\delta$ are complex numbers, $\vert a\rangle,\vert b\rangle,\vert c\rangle,$
$\vert d\rangle$ represent normalized single-mode states with non-zero
inner products $\langle a\vert c\rangle$ and $\langle b\vert d\rangle$,
and $M$ denotes the normalization constant. The concurrence is defined
as \citep{PhysRevLett.80.2245,PhysRevA.68.043606}:
\begin{equation}
C=\left|2M^{2}\lambda\delta\sqrt{1-\vert P_{1}\vert^{2}}\sqrt{1-\vert P_{2}\vert^{2}}\right|,\label{eq:23}
\end{equation}
 where $P_{1}$ and $P_{2}$ are given by $P_{1}=\langle a\vert c\rangle$
and $P_{2}=\langle b\vert d\rangle$. This quantity ranges from $0$
(separable) to $1$ (maximally entangled). Mapping our Bell-like states
from Eqs. (\ref{eq:24-1}) and (\ref{eq:26}) to this general form
yields the concurrence expression: 
\begin{equation}
C=\left|\frac{\left(1-e^{-2s^{2}}\right)\left(1-\vert\langle A\rangle_{\omega}\vert^{2}\right)}{1+e^{-2s^{2}}+\vert\langle A\rangle_{w}\vert^{2}\left(1-e^{-2s^{2}}\right)}\right|.\label{eq:29}
\end{equation}
This expression applies equally to $\vert\Phi_{3}\rangle$ and $\vert\Phi_{4}\rangle$.
Notably, due to the properties of coherent states, the concurrence
depends exclusively on the coupling strength $s$ and weak value $\langle A\rangle_{w}$,
independent of the input coherent amplitudes. The concurrence vanishes
when $s=0$ or $\langle A\rangle_{w}=\pm1$, while approximately reaching
its maximum value ($C=1$) under strong interaction conditions with
null and anomalously large weak values. This demonstrates the essential
role of postselected von Neumann measurements in generating entanglement
from initially uncorrelated states. The numerical analysis of concurrence
for our Bell-like states is presented in Figure \ref{fig:7}.

Figure \ref{fig:7} (a) illustrates the concurrence $C$ as a function
of both the weak value $\langle A\rangle_{w}$ and the interaction
strength $s$. The concurrence increases monotonically with $s$,
approximately reaching its maximum value of $1$ for sufficiently
large anomalous weak values. Notably, even for very small or zero
weak values, a substantially large $s$ still yields maximally entangled
states in the coherent-state basis. These trends are further corroborated
by the two-dimensional cross-sections presented in Fig. \ref{fig:7}
(b) and (c). Based on both exact mathematical derivation and numerical
evidence, we conclude that our protocol enables the preparation of
maximally entangled Bell states--- $\vert\psi_{1}\rangle$, $\vert\psi_{2}\rangle$,
$\vert\psi_{3}\rangle$ and $\vert\psi_{4}\rangle$---with appreciable
success probabilities.

Our feasible nG state preparation protocol provides a versatile method
for generating a wide range of two-mode entangled states. The degree
of quantum correlation in these states can be systematically enhanced
by tuning relevant system parameters. Crucially, the postselected
von Neumann measurement---governed by the choice of postselection
and the resulting weak value---plays an indispensable role in this
entanglement enhancement process. We also note that our protocol for
generating various two-mode entangled states can be implemented in
optical systems that incorporate postselected weak measurement techniques. 

\section{\label{sec:5}conclusion and outlook }

This study has introduced a feasible protocol for preparing nG quantum
states utilizing postselected von Neumann measurements. We have elucidated
the fundamental principles of this measurement technique and the role
of weak values in generating diverse single- and two-mode nG states,
including intermediate states arising from various pointer states
with non-negative Wigner distributions. Notably, our protocol achieves
substantially higher success probabilities compared to conventional
nG operation-based methods.

Specifically, we demonstrated that inputting squeezed and coherent
states enables the generation of GKP-like and cat-like states, respectively,
with high success probabilities in the single-mode case. For two-mode
scenario of our protocol, using uncorrelated Gaussian states as inputs
and applying postselected von Neumann measurements to one mode followed
by a BS operation yield various entangled states---including (squeezing)
two-mode entangled cat states, maximally entangled Bell states, and
continuous families of intermediate states---with considerable success
rates. The quality of these generated states was verified through
joint Wigner function analysis and quantified using entanglement measures
such as linear entropy and concurrence for CV systems.

In the present work, we have focused on the case where the pointer
variable is the momentum operator ($Q=P$). The protocol can be extended
to the position operator case, where the von Neumann interaction Hamiltonian
$H_{x}=g_{0}\sigma_{x}(a+a^{\dagger})$ can be realized in quantum
optics platforms using the Jaynes-Cummings model \citep{Agarwal2013}.
For $Q=X$ the output state becomes $\left[t_{+}D(-ig\sigma)+t_{-}D(ig\sigma)\right]\vert\phi\rangle$
(unnormalized), enabling the preparation of cat-like states applicable
to quantum random walks \citep{PhysRevA.68.020301,PhysRevA.75.032351}
in quantum electrodynamics systems.

Our results demonstrate the significant potential of postselected
von Neumann measurements for quantum state engineering and provide
a promising approach for generating high-quality nG states. Given
the extensibility of our protocol to multi-mode scenarios, future
work could explore its application for preparing advanced multiphoton
entangled states such as GHZ \citep{Greenberger1989} and NOON states
\citep{PhysRevA.40.2417,1188172}. 

Furthermore, while our analysis demonstrates the high efficiency of
this protocol under ideal conditions, the robustness of the generated
nG states against decoherence constitutes a vital area for future
investigation. The highly non-classical features, such as Wigner negativity
and entanglement, are particularly susceptible to photon loss and
thermal noise. A detailed study of the evolution of these states in
lossy channels and their performance in specific quantum protocols,
such as quantum metrology or error correction, will be crucial for
assessing their practical utility and guiding experimental implementations.
\begin{acknowledgments}
This work was supported by the National Natural Science Foundation
of China (No. 12365005).

\end{acknowledgments}
\bibliographystyle{apsrev4-1}
\bibliography{Reference}

\end{document}